\documentclass[11pt]{article}
\usepackage{amssymb}
\usepackage{amsmath}
\usepackage{amsthm}

\usepackage{graphicx,color,colordvi}

\def\openone{\leavevmode\hbox{\small1\kern-3.8pt\normalsize1}}

\newtheorem{theorem}{Theorem}
\newtheorem{lemma}{Lemma}
\newtheorem{proposition}{Proposition}

\newtheorem{conjecture}{Conjecture}
\theoremstyle{definition}

\newcommand{\lge}{\left\{\begin{array}{c}\le\\ \ge\end{array}\right\}}
\newcommand{\gle}{\left\{\begin{array}{c}\ge\\ \le\end{array}\right\}}

\newcommand{\half}{\mbox{$\textstyle \frac{1}{2}$} }

\def\reff#1{(\ref{#1})}

\newcommand{\twomat}[4]{{\left(\begin{array}{cc}#1&#2\\#3&#4\end{array}\right)}}
\newcommand{\supp}{\mathop{\rm supp}\nolimits}
\newcommand{\trace}{\mathop{\rm Tr}\nolimits}
\newcommand{\tr}{\mathop{\rm Tr}\nolimits}
\newcommand{\re}{\mathop{\rm Re}\nolimits}
\newcommand{\im}{\mathop{\rm Im}\nolimits}
\newcommand{\Sp}{\mathop{\rm Sp}\nolimits}

\newcommand{\diag}{\mathop{\rm diag}\nolimits}

\newcommand{\cD}{{\cal D}}

\newcommand{\cH}{{\cal H}}
\newcommand{\cI}{{\cal I}}

\newcommand{\cP}{{\cal P}}
\newcommand{\cS}{{\cal S}}

\newcommand{\cR}{{\cal R}}

\newcommand{\C}{{\mathbb{C}}}
\newcommand{\R}{{\mathbb{R}}}

\newcommand{\M}{{\mathbb{M}}}

\DeclareRobustCommand\openone{\leavevmode\hbox{\small1\normalsize\kern-.33em1}}
\newcommand{\identity}{I}%{\mathrm{\openone}}
\newcommand{\id}{\identity}
\newcommand{\be}{\begin{equation}}
\newcommand{\ee}{\end{equation}}
\newcommand{\bea}{\begin{eqnarray}}
\newcommand{\eea}{\end{eqnarray}}
\newcommand{\beas}{\begin{eqnarray*}}
\newcommand{\eeas}{\end{eqnarray*}}
\def\cDH{\cD(\cH)}
\def\cPH{\cP(\cH)}

\begin{document}
\title{\LARGE\bf
  $\alpha$-$z$-R\'enyi relative entropies}
\author{Koenraad M.R.~Audenaert$^{1,2}$ and Nilanjana Datta$^3$\\[3mm]
  \small\it $^1$Department of Mathematics, Royal Holloway University of London, Egham TW20 0EX, U.K.\\[1mm]
  \small\it $^2$Department of Physics and Astronomy, Ghent University, \\
  \small\it S9, Krijgslaan 281, B-9000 Ghent, Belgium\\[1mm]
  \small\it $^3$Statistical Laboratory, Centre for Mathematical Sciences, University of Cambridge,\\
  \small\it Wilberforce Road, Cambridge CB3 0WA, U.K.\\[1mm]
}

\date{}

\maketitle

\begin{abstract}
We consider a two-parameter family of  R\'enyi relative entropies $D_{\alpha,z}(\rho||\sigma)$
that are quantum generalisations of the classical R\'enyi divergence $D_{\alpha}(p||q)$.
This family includes many known relative entropies (or divergences) such as the quantum relative entropy,
the recently defined quantum R\'enyi divergences, as well as the quantum R\'enyi relative entropies.
All its members satisfy the quantum generalizations of R\'enyi's axioms for a divergence.
We consider the range of the parameters $\alpha,z$ for which the data processing inequality holds.
We also investigate a variety of limiting cases for the two parameters, obtaining explicit formulas for each one of them.
\end{abstract}

\section{Introduction}\label{intro}
The {\em{quantum relative entropy}} as introduced by Umegaki \cite{umegaki} is the proper \cite{hiaipetz91}
quantum generalisation of the classical Kullback-Leibler divergence
and it therefore plays a central role in quantum information theory.
In particular, fundamental limits on the performance of information-processing tasks
in the so-called ``asymptotic, memoryless (or i.i.d.) setting''
is given in terms of quantities derived from the quantum relative entropy.

There are, however, several other entropic quantities and generalized relative entropies (or divergences) which are
also of operational significance. One of the most important of these is the
family of relative entropies called the \emph{$\alpha$-R\'enyi relative entropies}
($\alpha$-RRE)
$D_\alpha(\rho||\sigma)$, where $\alpha \in (0,1) \bigcup (1,\infty)$,
which are quantum generalisations of the classical R\'enyi divergences.
For $\alpha \in (0,1)$  these relative entropies arise in the quantum Chernoff bound
\cite{chernoff}
which characterizes the probability of error in discriminating two different quantum states in the setting of asymptotically many copies.
In analogy with the operational interpretation of their classical counterparts, the $\alpha$-RRE can be viewed
as generalized cutoff rates in quantum binary state discrimination~\cite{milan}.

%Another important class of entropies are the smooth min- and max-entropies~\cite{renner} and the generalized relative entropies
%(namely the {\em{min- and max-relative entropies}})
%from which they are derived~\cite{dupuis, min-max}, which play a pivotal role in ``one-shot quantum information theory'',
%in which the assumption of an asymptotic i.i.d.\ setting is lifted
%(see e.g.~\cite{marco} and references therein).
%
%The list of entropic quantities still goes on,
%and includes the quasi-entropy~\cite{petz}, skew divergence~\cite{skew}, Tsallis entropy~\cite{tsallis} and subentropy~\cite{sub}
%to name a few.
In the light of this plethora of different entropic quantities that arise in quantum information theory,
it is desirable to find a mathematical framework that unifies as many of these quantities as possible.
Recently, a non-commutative generalization of the $\alpha$-RRE was defined that partially provided
such a framework.
Known alternatively as the {\em{$\alpha$ quantum R\'enyi divergence}} ($\alpha$-QRD) or the
\emph{``sandwiched''   R\'enyi relative entropy}, it depends
on a parameter $\alpha \in (0,1) \bigcup (1,\infty)$~\cite{marcotalk,fehrtalk,WWY, ML}.
For two positive semidefinite operators $\rho$ and $\sigma$ we denote it as
$\widetilde D_\alpha(\rho||\sigma)$. It has been proved to reduce to the min-relative entropy when $\alpha=1/2$, to
the quantum relative entropy in the limit $\alpha \to 1$, and to the max-relative entropy in the limit $\alpha \to \infty$ \cite{dupuis, min-max}.
Consequently, many properties of the min-, max- and quantum relative entropies can be inferred directly from those of
the $\alpha$-QRD. For example, the data-processing inequality (i.e.\ monotonicity under
completely positive trace-preserving maps) of these relative entropies is implied by that of $\widetilde D_\alpha(\rho||\sigma)$
for $\alpha \ge 1/2$~\cite{FL, Beigi}. The fact that the min- and max-relative entropies provide lower and upper bounds
to  the quantum relative entropy follows directly from the fact that the function $\widetilde D_\alpha(\rho||\sigma)$ is monotonically increasing in
$\alpha$~\cite{ML}. Also joint convexity of the min- and quantum relative entropies is implied by the joint convexity of
$\widetilde D_\alpha(\rho||\sigma)$ for $1/2 \le \alpha \le 1$ \cite{FL}.
%Furthermore, an operational interpretation of $\widetilde D_\alpha(\rho||\sigma)$
%was provided in \cite{milan2} in the context of quantum hypothesis testing.

In spite of these and various other interesting properties, which have been proved using a variety of sophisticated mathematical
tools, the framework of the $\alpha$-QRD family has certain limitations: $(i)$ the data-processing inequality, which is one of the
most desirable properties of any divergence-type quantity, is not satisfied for $\alpha \in (0,1/2)$~\cite{ML, DL10}, and
$(ii)$ the $\alpha$-QRD family is not the only quantum generalisation of the classical R\'enyi divergences,
as it does not incorporate the previously mentioned $\alpha$-RRE family.

In this paper we address both limitations by introducing a two-parameter family of quantum relative entropies that generalise
the classical R\'enyi divergences. We refer
to them as {\em{$\alpha$-$z$-R\'enyi relative entropies}} ($\alpha$-$z$-RRE), and denote them as $D_{\alpha, z}(\rho||\sigma)$,
with $\alpha$ and $z$ being two real parameters. For every value of the parameter $z$ one thus obtains a different, continuously varying
quantum generalisation of $D_\alpha(p||q)$.
This new family satisfies the data processing inequality (DPI) for \emph{all} values of $\alpha$, with certain restrictions on the
parameter $z$ as indicated below.
Furthermore, both the $\alpha$-QRD and the $\alpha$-RRE are included as special cases (for $z=\alpha$ and $z=1$, respectively).

In Section~\ref{main} we  define this new family of relative entropies and summarize our main results. We state how the other known relative
entropies can be obtained from this family; we prove that the $\alpha$-$z$-RRE satisfies
the quantum generalizations of R\'enyi's axioms for a divergence, and describe
the regions in the $\alpha$-$z$ plane where these entropies satisfy  the data-processing inequality.
We study a special case of the $\alpha$-$z$-RRE, which we denote as  $\widehat D_\alpha$ (and informally call the
\emph{reverse sandwiched R\'enyi relative entropy})
due to its similarities with the $\alpha$-QRD (or sandwiched R\'enyi relative entropy).
It satisfies the data-processing inequality for $\alpha \le 1/2$, and we
obtain an interesting closed expression for it in the limit $\alpha \to 1$.
In Sections~\ref{sec:limit0}, \ref{sec:limitinf} and \ref{sec:lim00}
we study limiting cases of the $\alpha$-$z$-RRE.
We end the paper with a brief summary of our results and some open questions in Section~\ref{discuss}.

Obtaining a single quantum generalization of the classical R\'enyi divergence, which would cover all possible operational scenarios in
quantum information theory, is a challenging (and perhaps impossible) task. However, we believe that the $\alpha$-$z$-RRE is currently the best
candidate for such a quantity, since it unifies all known quantum relative entropies in the literature to date.

\section{Definitions and Main Results}\label{main}
Throughout the paper $\cH$ denotes a finite-dimensional Hilbert space.
We denote by $\cPH$ the set of positive semidefinite operators on $\cH$ and by $\cDH$ the set of density operators on $\cH$, i.e.\ operators
$\rho\in\cPH$ with $\tr\rho=1$. Further, we denote the support of an operator $\rho$ by $\supp\rho$.
Logarithms are taken to base $2$.
We denote the ordered eigenvalues of a $d\times d$ Hermitian matrix $X$ as
$\lambda_1(X)\ge \lambda_2(X)\ge\ldots\ge \lambda_d(X)$.

Let us first give the definition of the $\alpha$-$z$-R\'enyi relative ($\alpha$-$z$-RRE) entropies;
$\forall \rho\in \cD(\cH), \sigma\in\cP(\cH)$ with $\supp\rho\subseteq\supp\sigma$
\be
D_{\alpha,z}(\rho||\sigma) :=
\frac{1}{\alpha-1}\log f_{\alpha,z}(\rho||\sigma),
\ee
where $f_{\alpha,z}(\rho||\sigma)$ is the trace functional
\bea
f_{\alpha,z}(\rho||\sigma)
&:=& \trace\left(\rho^{\alpha/2z} \sigma^{(1-\alpha)/z}\rho^{\alpha/2z}\right)^{z} \label{eq:frs2} \\
&=& \trace\left(\sigma^{(1-\alpha)/2z}\rho^{\alpha/z} \sigma^{(1-\alpha)/2z}\right)^{z}.\label{eq:frs3}
\eea
Here, $\alpha\in\R$ and the limit has to be taken for $\alpha$ tending to 1, and $z\in\R^+$ and the limit has to be taken for $z$ tending to 0.
Also, negative powers are defined in the sense of generalized inverses; that is, for negative $x$, $\rho^x:=(\rho|_{\supp\rho})^x$.
The above definition is easily extended to the case in which $\rho \ge 0$ but $\tr \rho \ne 1$ (see \reff{def_gen}).
The trace functional can be written alternatively as
\bea
f_{\alpha,z}(\rho||\sigma)
&=& \trace\left(\rho^{\alpha/z} \sigma^{(1-\alpha)/z}\right)^{z}. \label{eq:frs1}
\eea
This is because for any pair of square matrices $A$ and $B$, the eigenvalues of $AB$ and $BA$ are the same (see, e.g.\ \cite{bhatia}, exercise I.3.7).
Hence, the matrix $\rho^{\alpha/z} \sigma^{(1-\alpha)/z}$
has real, non-negative eigenvalues (even though it is not in general self-adjoint),
and the trace functional $\trace(\cdot)^z$ in this expression is well-defined as the sum of $z$th powers of these eigenvalues,
which are the same as those of $\rho^{\alpha/2z} \sigma^{(1-\alpha)/z}\rho^{\alpha/2z}$.

For commuting $\rho$ and $\sigma$, $D_{\alpha,z}(\rho||\sigma)$ reduces to the classical $\alpha$-R\'enyi divergence,
for all values of $z$, as required.

Clearly, this family includes the $\alpha$-RRE family:
\be
D_\alpha(\rho||\sigma) := \frac{1}{\alpha-1}\log \trace\left(\rho^{\alpha} \sigma^{1-\alpha}\right) = D_{\alpha,1}(\rho||\sigma),
\ee
and the $\alpha$-QRD family:
\be
\widetilde D_\alpha(\rho||\sigma) :=
\frac{1}{\alpha-1}\log \trace\left(\sigma^{\frac{1-\alpha}{2\alpha}} \rho \sigma^{\frac{1-\alpha}{2\alpha}}\right)^{\alpha}
= D_{\alpha,\alpha}(\rho||\sigma).
\ee
Specifically, we get the known correspondences \cite{ML}
\be
D_{\min} = D_{1/2,1/2}, \quad D = \lim_{\alpha\to 1}D_{\alpha,\alpha}, \quad \mbox{and } D_{\max} = \lim_{\alpha\to \infty}D_{\alpha,\alpha}.
\ee
Here $D_{\min}$, $D$ and $D_{\max}$ denote the min-relative entropy~\cite{dupuis},
the quantum relative entropy and the max-relative entropy~\cite{min-max}, respectively:
\bea
D_{\min}(\rho||\sigma) &:=& - 2 \log F(\rho, \sigma), \quad {\hbox{where}} \quad F(\rho, \sigma)= ||\sqrt{\rho}\sqrt{\sigma}||_1,\nonumber\\
D(\rho||\sigma) &:=& \tr \rho \log \rho - \tr \rho \log \sigma,\nonumber\\
D_{\max}(\rho||\sigma) &:=& \inf \{\gamma : \rho \le 2^\gamma \sigma\}.
\label{dmax2}
\eea

\begin{figure}[ht]
\begin{center}
\includegraphics[width=14cm]{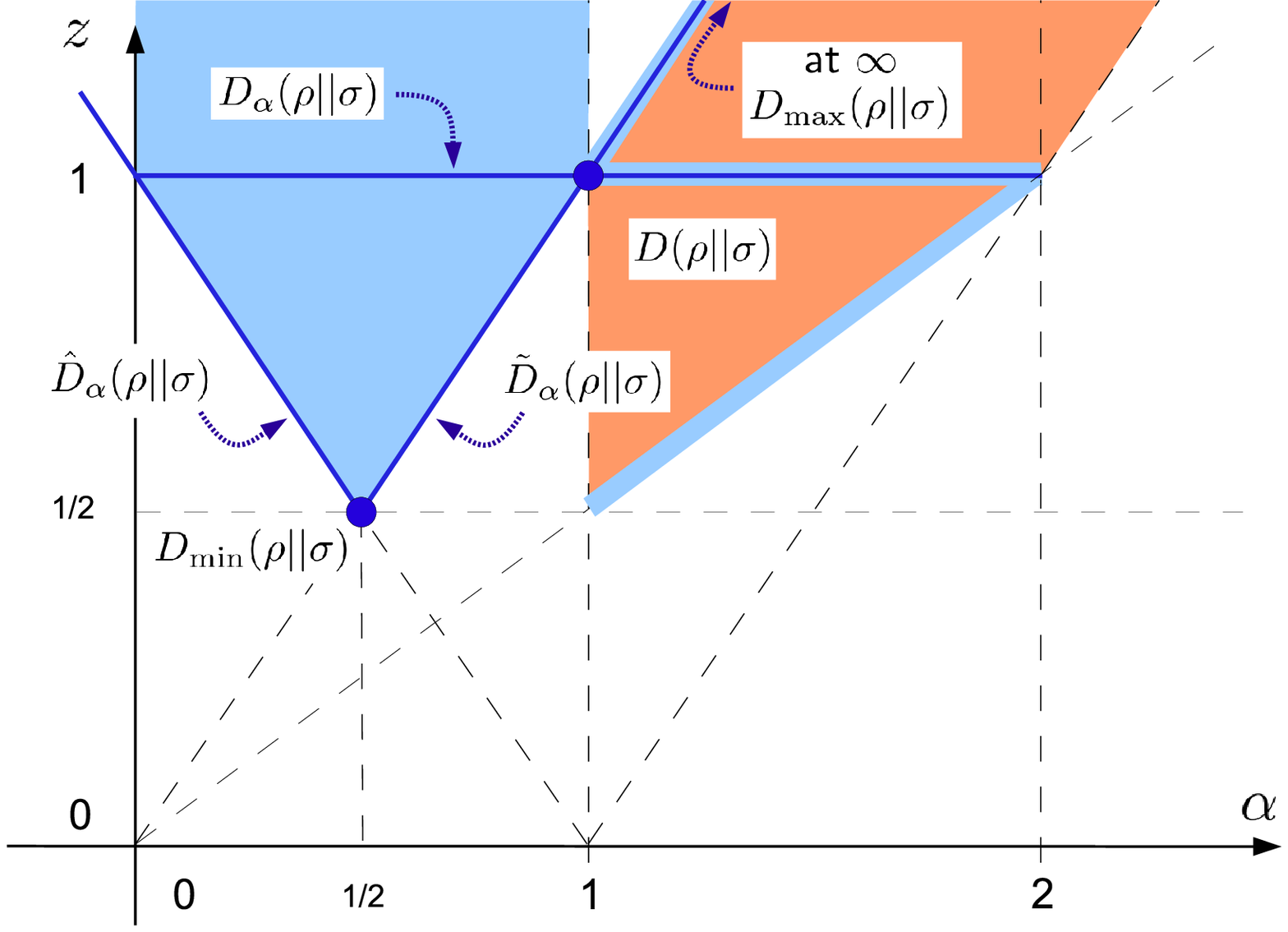}
\end{center}
\caption{Schematic overview of the relative entropies that are unified by $D_{\alpha,z}$, as indicated
by the dark-blue lines and dots.
The region where the Data Processing Inequality (DPI)
has been proven to hold has been coloured light-blue,
and the orange region is where we conjecture validity of DPI.
Outside these two regions DPI does not hold.
For details, see Section \ref{sec:DPI}.}
\end{figure}

These correspondences are illustrated in Figure~1.
Also included in the family is a quantity defined by Hayashi in \cite{hayashi}, which essentially is $D_{\alpha,2}$.
Furthermore, as was pointed out by Lin and Tomamichel \cite{lintoma},
the derivatives of $D_{\alpha,1}$ and $D_{\alpha,\alpha}$ with respect to $\alpha$ and taken at $\alpha=1$ are both equal to one half the so-called
\emph{quantum information variance} \cite{REV1,REV2}
\[
V(\rho||\sigma):=\trace\rho(\log\rho-\log\sigma)^2 - D(\rho||\sigma)^2.
\]

The epithet ``sandwiched'' in the original name of the $\alpha$-QRD stems from the fact that in its formula $\rho$
appears sandwiched between two powers of $\sigma$.
Now note that one could also consider another way of sandwiching by putting $\sigma$ between two powers of $\rho$,
modifying the exponents accordingly so that the functional again coincides with $D_\alpha$ in the commutative setting.
This new quantity $\widehat D_\alpha$ (which we informally call the \emph{reverse sandwiched R\'enyi relative entropy}) is defined as
\be
\widehat D_\alpha(\rho||\sigma)
= \frac{1}{\alpha-1}\log \trace\left(\rho^{\frac{\alpha}{2(1-\alpha)}} \sigma \rho^{\frac{\alpha}{2(1-\alpha)}}\right)^{1-\alpha}
= D_{\alpha,1-\alpha}(\rho||\sigma).
\ee
From (\ref{eq:symm}) we immediately obtain the symmetry relation
\be
(\alpha-1)\widehat D_\alpha(\rho||\sigma) = (-\alpha)\widetilde D_{1-\alpha}(\sigma||\rho).\label{eq:hatsymm}
\ee
For $\alpha=0$, $\widehat D_\alpha$ reduces to the 0-R\'enyi relative entropy, a quantity of particular operational
relevance in one-shot information theory \cite{wang,BD}. This is in contrast to the $\alpha$-quantum R\'enyi divergence,
which does not in general reduce to the 0-R\'enyi relative entropy in the limit $\alpha\to0$ \cite{DL10}.

\bigskip

\noindent\textbf{Remarks.}
\begin{enumerate}
\item For states $\rho$ and $\sigma$ with identical support, $D_{\alpha,z}$ is even in $z$: $D_{\alpha,z}(\rho||\sigma) = D_{\alpha,-z}(\rho||\sigma)$.
This is no longer the case when the support of $\rho$ is a proper subset of $\supp\sigma$. For example,
one easily checks that
$f_{2,-1}(\rho||\sigma) = \trace\rho^2(\sigma|_{\supp\rho})^{-1}$, whereas
$f_{2,1}(\rho||\sigma) = \trace\rho^2(\sigma^{-1})|_{\supp\rho}$.
Taking $z<0$ might therefore complicate matters substantially, whereas there is no guarantee that the results will be interesting. We therefore have limited
our considerations to $z>0$ throughout.
\item
The family obeys a symmetry condition with respect to $\alpha$:
\be
(\alpha-1) D_{\alpha,z}(\rho||\sigma) = (-\alpha) D_{1-\alpha,z}(\sigma||\rho). \label{eq:symm}
\ee

\item
The family coincides with certain quantum entropic functionals defined by
Jak\v{s}i\'c {\em{et al.}}~\cite{jaksic} for the study of entropic fluctuations in non-equilibrium quantum statistical mechanics.
These functionals were defined in the context of a dynamical system: in particular, $\rho$ was the reference state of a dynamical system,
and $\sigma$ was the state $\rho_t$ resulting from $\rho$ due to time evolution under the action of a Hamiltonian for a time $t$.
In contrast, we define $D_{\alpha, z}(\rho||\sigma)$ for arbitrary positive semidefinite states $\rho$ and $\sigma$, and study its properties from a
quantum information theoretic perspective.
\end{enumerate}
%%%%%%%%%%%%%%%%%%%%%%%%%%%%%%%%%%%%%%%%%%%%%%
%%%%%%%%%%%%%%%%%%%%%%%%%%%%%%%%%%%%%%%%%%%%%%
\subsection{Axiomatic properties}
Following \cite{ML}, we can check whether the $\alpha$-$z$-RRE satisfies the six quantum R\'enyi axioms, as do the $\alpha$-RRE and $\alpha$-QRD.
These are quantum generalizations of axioms that were put forward by R\'enyi in \cite{renyiaxioms} as natural requirements that any classical
divergence should satisfy. A quantum divergence is a functional which maps a pair of positive semidefinite operators
$\rho, \sigma$, with $\supp \rho \subseteq \supp \sigma$ onto
$\R$. Its classical counterpart is obtained by replacing the operators by probability distributions.

Within this context we need to slightly redefine the $\alpha$-$z$-RRE for non-normalized states $\rho$:
$\forall \rho,\sigma\in \cP(\cH)$ with $\supp\rho\subseteq\supp\sigma$,
\be\label{def_gen}
D_{\alpha,z}(\rho||\sigma) := \frac{1}{\alpha-1} \log \frac{f_{\alpha,z}(\rho,\sigma)}{\trace \rho}.
\ee

\begin{enumerate}
%%%%%%
\item[(I)] \textbf{Continuity:} For $\rho\neq0$ and $\supp\rho\subseteq\supp\sigma$,
$D_{\alpha,z}(\rho||\sigma)$ is continuous in $\rho,\sigma\ge0$ throughout the parameter space except for $\alpha\le 0$.
At $\alpha=0$, the $\alpha$-RRE is dependent on the rank of $\rho$ and is therefore not continuous.
This was actually the reason why R\'enyi included the continuity axiom: to exclude the cases $\alpha\le0$, where the relative entropy functional
was not deemed a reasonable measure of information (\cite{renyiaxioms}, p.\ 558) due to its discontinuity.

The only case where it is not obvious that continuity holds for $\alpha>0$ is the case $z=0$. This will be considered in Section \ref{sec:limit0}.
%%%%%%
\item[(II)] \textbf{Unitary invariance:} For unitary $U$, $D_{\alpha,z}(U\rho U^*||U\sigma U^*) = D_{\alpha,z}(\rho||\sigma)$.
%%%%%%
\item[(III)] \textbf{Normalization:} $D_{\alpha,z}(1||\half)=1$ (for scalar arguments, and when using base-2 logarithms),
as is the case for any divergence that reduces to the classical
R\'enyi divergence for commuting arguments.
%%%%%%
\item[(IV)] \textbf{Order Axiom:}
The axiom requires that
$$
D_{\alpha,z}(\rho||\sigma)\lge0 \mbox{ whenever } \rho\lge\sigma.
$$
Note that this axiom is a weaker version of the Data Processing Inequality (DPI) considered below, as follows from Lemma 5 in \cite{coles}.

\begin{proposition}
$D_{\alpha,z}$ satisfies the Order Axiom when $z\ge |\alpha-1|$.
\end{proposition}
\textit{Proof.}
Noting that $\trace\rho = f_{\alpha,z}(\rho||\rho)$,
we need, for $\alpha>1$,
$$
f_{\alpha,z}(\rho||\sigma) \lge  f_{\alpha,z}(\rho||\rho) \mbox{ whenever } \rho\lge\sigma,
$$
whereas, for $0<\alpha<1$,
$$
f_{\alpha,z}(\rho||\sigma) \lge  f_{\alpha,z}(\rho||\rho) \mbox{ whenever } \rho\gle\sigma.
$$
This holds if the fractional power $(1-\alpha)/z$ that is applied to $\sigma$ in (\ref{eq:frs2}) is operator monotone, when $0<\alpha<1$,
and operator monotone decreasing, when $\alpha>1$.
In other words, for $0<\alpha<1$, $(1-\alpha)/z$ must lie between $0$ and $1$, i.e.\ $z\ge (1-\alpha)$.
For $\alpha>1$ it must lie between $-1$ and $0$, i.e.\ $z\ge (\alpha-1)$.
\qed

In Figure~1 this corresponds to the triangular region with apex $(1,0)$ and
sides passing through the points $(0,1)$ and $(2,1)$, respectively.
%%%%%%
\item[(V)] \textbf{Additivity} with respect to tensor products: clearly,
$$
D_{\alpha,z}(\rho\otimes\tau||\sigma\otimes\omega) = D_{\alpha,z}(\rho||\sigma) + D_{\alpha,z}(\tau||\omega).
$$
%%%%%%
\item[(VI)] \textbf{Generalized Mean Value Axiom:}
This axiom describes the behavior of $D_{\alpha,z}$ with respect to direct sums (the quantum generalization of taking the union of incomplete
probability distributions). It requires
the existence of a continuous, strictly increasing function $g$
such that
$$
(\trace\rho+\trace\tau)\;g(D_{\alpha,z}(\rho\oplus \tau || \sigma\oplus\omega))
= (\trace\rho) \;g(D_{\alpha,z}(\rho||\sigma)) + (\trace\tau) \;g(D_{\alpha,z}(\tau||\omega)).
$$
In the classical case, if $g$ is affine this requires that the divergence between pairs of unions of distributions is a weighted arithmetic mean of
divergences, and this (along with the other axioms) limits $D$ to be the classical relative entropy.
Taking exponential $g$, $g(x)=\exp((\alpha-1)x)$, we obtain the classical R\'enyi divergences.

Now, to see that $D_{\alpha,z}$ satisfies this axiom, it is sufficient to note that
$$
f_{\alpha,z}(\rho\oplus \tau|| \sigma\oplus\omega)
= f_{\alpha,z}(\rho||\sigma) + f_{\alpha,z}(\tau||\omega).
$$
This holds throughout the parameter space, provided we choose $g(x) = \exp((\alpha-1)x)$, of course.

Note that in \cite{renyiaxioms} only the case $\trace \rho+\trace \tau\le 1$ and $\trace\sigma+\trace\omega\le 1$ is considered,
so that $\rho\oplus\tau$ and $\sigma\oplus\omega$  are normalized or subnormalized density matrices,
the quantum generalization of generalized (i.e.\ complete or incomplete)
probability distributions, but it turns out that even without this restriction the equality of the axiom holds.
\end{enumerate}
%%%%%%%%%%%%%%%%%%%%%%%%%%%%%%%%%%%%%%%%%%%%%%
%%%%%%%%%%%%%%%%%%%%%%%%%%%%%%%%%%%%%%%%%%%%%%
\subsection{Data Processing Inequality\label{sec:DPI}}
A more difficult question is for which parameter range $D_{\alpha,z}$ satisfies the Data Processing Inequality (DPI).
While this has not yet been established in full generality, it can be shown to hold for certain parameter ranges, indicated on Figure~1
by light-blue shading.
\begin{theorem}[Data-processing Inequality]
For any pair of positive semidefinite operators $\rho, \sigma \in \cPH$, for which $\supp\rho \subseteq \supp\sigma$, and
for any CPTP map $\Lambda$ acting on $\cPH$, the Data Processing Inequality
$$
D_{\alpha, z}\left(\Lambda(\rho)||\Lambda(\sigma) \right) \le D_{\alpha, z}\left(\rho||\sigma \right),
$$
holds in each of the following cases:
\begin{itemize}
\item $0<\alpha\le 1$ and $z \ge \max \left(\alpha,1-\alpha \right)$ (Hiai),
\item $1\le \alpha\le 2$ and $z=1$ (Ando),
\item $1\le \alpha$ and $z=\alpha$ (Frank and Lieb; Beigi),
\item $1\le\alpha\le2$ and $z=\alpha/2$ (Carlen, Frank and Lieb).
\end{itemize}
\end{theorem}
It is well-known that to prove DPI for $D_{\alpha,z}$
one has to show that the trace functional $f_{\alpha,z}(\rho||\sigma)$ that lies at the heart of $D_{\alpha,z}$ is
jointly concave when $\alpha\le1$, or
jointly convex when $\alpha\ge1$ (see, e.g. \cite{FL}, its \textit{Proof of Theorem 1 given Proposition 3}).
In fact, it suffices to show that the related trace functional $f_{\alpha,z}(A;K)$, defined as
\be
f_{\alpha,z}(A;K) := \trace(A^{\alpha/z} K A^{(1-\alpha)/z} K^*)^{1/z},
\ee
is concave/convex in $A$ (for any fixed matrix $K$) over the set of positive semidefinite matrices.
Joint concavity/convexity of the original functional $f_{\alpha,z}(\rho||\sigma)$
then follows by setting $K=\twomat{0}{\id}{0}{0}$ and $A=\rho\oplus\sigma$.

Concavity of $f_{\alpha,z}(A;K)$
in the case $0<\alpha\le 1$ and $z \ge \max \left(\alpha,1-\alpha \right)$  follows directly from a concavity theorem proven very recently
by Hiai \cite{hiai13} (see also the older work \cite{hiai01}), whose proof is based on the complex analysis techniques employed by Epstein in \cite{epstein}.
Note that this generalises Corollary 1.1 of \cite{lieb}.
Epstein's paper is rather terse and uses deep results from complex analysis.
A pedagogical introduction can be found, for example,
in the appendix of \cite{ruskai}.
In Section 6 we provide a detailed proof using
similar techniques as Epstein's, but more elementary
and tailored to the problem at hand.

Convexity was proven by Frank and Lieb \cite{FL} and independently by Beigi \cite{Beigi}
for the case $1\le \alpha$ and $z=\alpha$, where $D_{\alpha,z}$ reduces to
the QRD $\widetilde D_\alpha$.
Convexity for $1\le \alpha\le 2$ and $z=1$ is exactly Ando's theorem \cite{ando}.
Finally, after the appearance of the first version of this paper, Carlen, Frank and Lieb were able to prove DPI
in the case $1\le\alpha\le2$ and $z=\alpha/2$ \cite{CFL14}.

Hiai \cite{hiai13} also provides necessary conditions for concavity/convexity. The regions in the parameter space where these conditions
are \emph{not} satisfied are indicated in Figure~1 as white space. About the remaining region, indicated in orange, nothing definitive is known
other than that the conditions for necessity are satisfied. For this region we conjecture that the trace
functional is convex, which would imply that DPI holds here as well.

\begin{figure}[ht]
\begin{center}
\includegraphics[width=14cm]{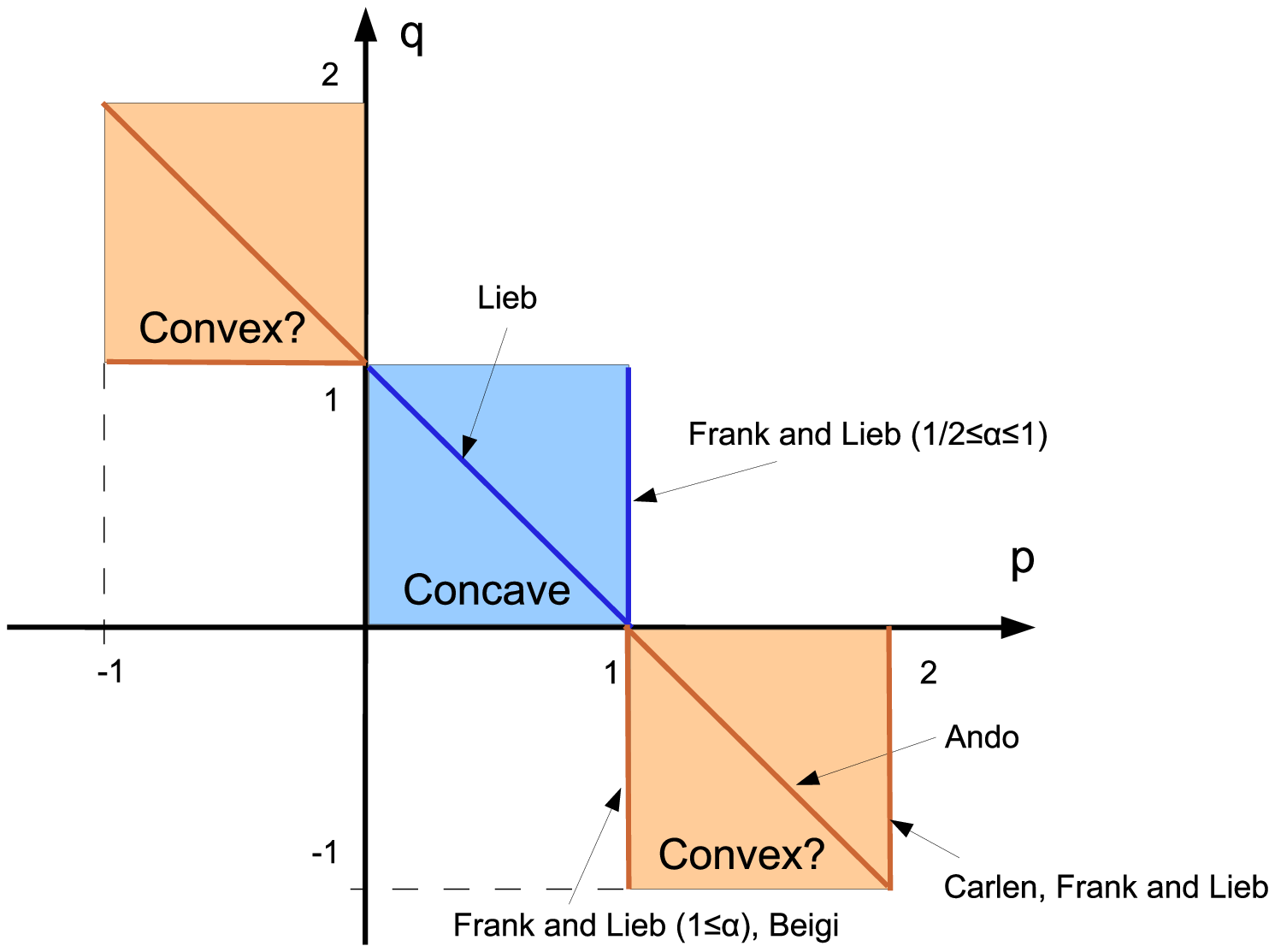}
\end{center}
\caption{Regions of concavity (blue, proven in this paper) and conjectured convexity (orange)
of the re-parameterized trace functional $f_{p,q}$, where $p=\alpha/z$ and
$q=(1-\alpha)/z$. The dark blue and dark orange lines indicate the values of $p$ and $q$ for which concavity and convexity have already been proven.
Note that the region where the Order Axiom is satisfied is the strip $-1\le q\le 1$ and excludes the upper left orange square.
The Continuity Axiom is satisfied in the region where $p$ and $p+q$ have the same sign ($\alpha>0$), again excluding the upper left
orange square.}\label{fig:2}
\end{figure}

When considering DPI, it is convenient to re-parameterize the trace functional $f_{\alpha,z}$  as
\be
f_{p,q}(A;K) := \trace(A^p K A^q K^*)^{1/(p+q)},
\ee
where the parameters $p$ and $q$ are defined as $p=\alpha/z$ and $q=(1-\alpha)/z$.
We obtain the original functional by setting $z=1/(p+q)$ and $\alpha=p/(p+q)$.
\begin{conjecture}\label{conj:1}
The trace functional $f_{p,q}(A;K)$ is convex on the set of positive definite $d\times d$ matrices for $-1\le p<0$ and $1\le q\le 2$ (or vice versa).
\end{conjecture}
Figure~\ref{fig:2} shows the regions in $(p,q)$-parameter space where DPI provably holds and where we conjecture it.
\medskip

\noindent\textbf{Remark.}
One notices that whereas the $\alpha$-QRD $\widetilde D_\alpha$ satisfies DPI only for $\alpha\ge 1/2$,
the reverse $\alpha$-QRD $\widehat D_\alpha$ satisfies DPI for $0\le\alpha\le 1/2$.

\subsection{Limiting cases}\label{sec:limit}
We study four limiting cases of the $\alpha$-$z$-RRE: (i) limit $\alpha\to1$ and $z\to 0$, (ii)
the case of infinite $\alpha$ and $z$, (iii) fixed $\alpha$ and infinite $z$, and (iv) $z=\alpha\to0$.

To study (i) we suitably parameterize $z$ in terms of $\alpha$ as $z=r(\alpha-1)$, where $r$ is a non-zero finite real number,
and consider the limit $\alpha \to 1$ (the case of fixed $\alpha\neq1$ and $z\to 0$ will be studied elsewhere \cite{KAHiai}).
Note that $\alpha=1$ is the only value of $\alpha$ where in the limit $z\to0$ the Order Axiom (IV) is satisfied.
For the choice $z=1-\alpha$, this yields the limit $\alpha \to 1$ of $\widehat{D}_\alpha(\rho||\sigma)$.
In the general case in which $\rho$ and $\sigma$ do not commute, we obtain a rather surprising formula for the latter:
the relative entropy, not between $\rho$ and $\sigma$, but between $\rho$ and an operator $\widehat{\sigma}$ that is diagonal in the eigenbasis
of $\rho$ (see Theorems~\ref{th:D10} and \ref{thm_3} for details).
In the commuting case we recover the expected expression:
the relative entropy of $\rho$ and $\sigma$. We also prove that the $\alpha$-$z$-RRE
is continuous in $\rho$ and $\sigma$ in that limit.

To study the case (ii) of infinite $\alpha$ and $z$, we use the same parametrization of $z$, and take the limit $\alpha \to \infty$. In this
limit the  $\alpha$-$z$-RRE is expressed in terms of a max-relative entropy (see Theorem~\ref{thm_4} for details).
In particular, our result readily yields the known \cite{ML} result that in the limit $\alpha \to \infty$,
the $\alpha$-QRD, $\widetilde{D}_{\alpha}(\rho||\sigma)$, reduces to the max-relative entropy $D_{\max}(\rho||\sigma)$.

Case (iii) concerns keeping $\alpha$ fixed (and finite) letting $z$ tend to $+\infty$. Using the Lie-Trotter relation,
we obtain the quantity $(1/(\alpha-1))\log\tr\exp(\alpha\log\rho+(1-\alpha)\log\sigma)$,
which in the limit $\alpha\to1$ tends to the relative entropy $D(\rho||\sigma)$.

Finally, we consider the case (iv) where $\alpha$ and $z$ both tend to 0, with $z=\alpha$.
%%%%%%%%%%%%%%%%%%%%%%%%%%%%%%%%%%%%%%%%%%%%%%%%%%%%%%%%%%%%%%%%%%%%%%%%%%%%%%%%%%%%%%%%%%%%%%%%%%%%%%%%%%%%%%%%%%%%%%%%%%%%%%%%%%%%%%%%%%%%%%
%%%%%%%%%%%%%%%%%%%%%%%%%%%%%%%%%%%%%%%%%%%%%%%%%%%%%%%%%%%%%%%%%%%%%%%%%%%%%%%%%%%%%%%%%%%%%%%%%%%%%%%%%%%%%%%%%%%%%%%%%%%%%%%%%%%%%%%%%%%%%%
\section{Limiting case $\alpha\to1$ and $z\to 0$}\label{sec:limit0}
In this section, we derive a closed form expression for the limit of $D_{\alpha,z}$ as $z$ tends to 0.
The most interesting point to calculate this is when $\alpha=1$ because that is the only value where the Order Axiom remains satisfied as $z$ goes to 0,
even though DPI no longer holds.
It turns out that $\lim_{z\to0}D_{\alpha,z}$ is discontinuous in $\alpha$ at $\alpha=1$
and we will have to be careful how the limit $z\to 0$ is taken.
What we will consider is the limit $\alpha\to 1$ of $D_{\alpha,r(\alpha-1)}$, with fixed $r$, i.e.\ the limit along straight lines
passing through the point $(1,0)$ and with slope $r$. This choice is particularly convenient since for $r=-1$ we recover the limit
$\lim_{\alpha\to 1} \widehat D_\alpha$.

As we assume $\supp\rho\subseteq\supp\sigma$ throughout, there is no loss of generality in only
considering $\sigma>0$; that is, all matrices will be restricted
to the subspace $\supp\sigma$.
\begin{lemma}
For $\sigma>0$, and $r$ a non-zero finite real number,
$$
\lim_{\alpha\to 1} \left(\rho^{\alpha/2r(\alpha-1)} \sigma^{-1/r} \rho^{\alpha/2r(\alpha-1)} \right)^{r(\alpha-1)} = \rho.
$$
\end{lemma}
\textit{Proof.}
Since $\sigma>0$, there exist $a,b>0$ such that $a\le\sigma\le b$ (meaning that $a\id\le\sigma\le b\id$).
Then, for $r>0$, $b^{-1/r}\le\sigma^{-1/r}\le a^{-1/r}$ so that
$$
b^{-1/r} \rho^{\alpha/r(\alpha-1)}
\le \rho^{\alpha/2r(\alpha-1)} \sigma^{-1/r} \rho^{\alpha/2r(\alpha-1)}
\le  a^{-1/r} \rho^{\alpha/r(\alpha-1)}.
$$
For $r<0$ the roles of $a$ and $b$ get interchanged.

Raising this to the power $r(\alpha-1)$, for $\alpha>1$ and close enough to 1 so that this is an operator monotone operation, yields
$$
b^{1-\alpha} \rho^{\alpha}
\le \left(\rho^{\alpha/2r(\alpha-1)} \sigma^{-1/r} \rho^{\alpha/2r(\alpha-1)}\right)^{r(\alpha-1)}
\le  a^{1-\alpha} \rho^{\alpha}.
$$
For $\alpha<1$ and close enough to 1, $a$ and $b$ again have to be interchanged (as it is an operator monotone decreasing operation).

In the limit $\alpha\to1$ we then get that $a^{1-\alpha}$ and $b^{1-\alpha}$ both tend to 1, and these inequalities become
$$
\rho
\le \lim_{\alpha\to 1}\left(\rho^{\alpha/2r(\alpha-1)} \sigma^{-1/r} \rho^{\alpha/2r(\alpha-1)}\right)^{r(\alpha-1)}
\le  \rho.
$$
As both bounds are equal, this proves that the inequalities actually are equalities.
\qed

A simple corollary of this lemma is that $\lim_{\alpha\to 1} f_{\alpha,r(\alpha-1)} = \trace\rho = 1$.
Hence, as $\alpha$ tends to $1$, both the numerator and denominator in $D_{\alpha,r(\alpha-1)} = \log f_{\alpha,r(\alpha-1)}/(\alpha-1)$
tend to $0$. To calculate the limit it is tempting to use l'H\^opital's rule and calculate
the derivative with respect to $\alpha$. However, this approach did not yield any simplification.
Instead, we followed a completely different approach, inspired by the power method \cite{golub} for numerically calculating eigenvalues.

We first consider the generic case in which the spectrum of $\rho$ is non-degenerate, i.e.\ all its eigenvalues are distinct.
Let us write the spectral decomposition of $\rho$ as
$\rho =\sum_{i=1}^d \mu_i P_i$, where the eigenvalues $\mu_i$ appear sorted in decreasing order and where $P_i$ are the corresponding
projectors $|i\rangle\langle i|$ on the (1-dimensional) eigenspaces.
The main idea behind the power method is that for large positive $s$,  $\rho^s$ can be well-approximated
by $\mu_1^s P_1$, in the sense that the sum of the remaining terms $\sum_{i=2}^d \mu_i^s P_i$ becomes much smaller in norm than $\mu_1^s$.

Let us denote the matrix expression inside the trace of the trace functional $f_{\alpha,r(\alpha-1)}$ by $Z_{\alpha,r}(\rho||\sigma)$.
Rather than applying the above approximation to the entire trace of $Z_{\alpha,r}(\rho||\sigma)$, which would be too crude,
we apply it to the calculation of its largest eigenvalue $\lambda_1$ only.
We get, for $z=r(\alpha-1)>0$,
\beas
\lambda_1 (Z_{\alpha,r}(\rho||\sigma)) &=& \lambda_1\left( (\rho^{\alpha/2r(\alpha-1)} \sigma^{-1/r} \rho^{\alpha/2r(\alpha-1)})^{r(\alpha-1)} \right) \\
&\approx& \mu_1^\alpha \trace(P_1 \sigma^{-1/r} P_1)^{r(\alpha-1)} \\
&=& \mu_1^\alpha \left((\sigma^{-1/r})_{1,1}\right)^{r(\alpha-1)},
\eeas
where $X_{1,1}$ indicates the upper left matrix element of a matrix $X$ in the eigenbasis of $\rho$.
This is shown in full rigor in Lemma \ref{lem:app} below.

As we ultimately need an expression for the trace we need approximations for all eigenvalues of $Z_{\alpha,r}$.
To proceed, we will use the so-called ``Weyl trick'', which consists in calculating the largest eigenvalue of the
$k$th antisymmetric tensor power of $Z_{\alpha,r}$
(see e.g.\ \cite{bhatia} Section I.5 for antisymmetric tensor powers and Section IX.2 for applications of the Weyl trick).
For any given matrix $X$, its $k$th antisymmetric tensor power, denoted $X^{\wedge k}$, is defined as the restriction of its
$k$th tensor power $X^{\otimes k}$ to the totally antisymmetric subspace. The reason for looking into this is that the largest eigenvalue
of $X^{\wedge k}$ is the product of the $k$ largest eigenvalues of $X$, an identity which we denote by the shorthand
$$
\lambda_1(X^{\wedge k}) = \lambda_1\cdots\lambda_k(X) := \lambda_1(X)\cdots\lambda_k(X).
$$
Furthermore, we have the relations $(XY)^{\wedge k} = X^{\wedge k} Y^{\wedge k}$ and $(X^s)^{\wedge k} = (X^{\wedge k})^s$.

For $X$ of dimension $d$, $k$ can take values from 1 to $d$. For $k=d$, the totally antisymmetric subspace is 1-dimensional and the antisymmetric
tensor power $X^{\wedge d}$ is a scalar, namely the determinant of $X$.
Analogously, the matrix elements of $X^{\wedge k}$ for $k<d$ are all possible $k\times k$ minors of $X$ (determinants of submatrices).
In particular, the ``upper left'' element $(X^{\wedge k})_{1,1}$ is the leading principal $k\times k$ minor of $X$.
If we introduce the notation $X_{1:k,1:k}$ to mean the submatrix of $X$ consisting of the first $k$ rows and the first $k$ columns,
this element is given by
$$
(X^{\wedge k})_{1,1} = \det\left(X_{1:k,1:k}\right).
$$

Let us now apply the power method to $Z_{\alpha,r}^{\wedge k}$ in order to obtain an approximation
for the product of the $k$ largest eigenvalues of $Z_{\alpha,r}$.
We will denote this product by $\lambda^{(k)}$, and by convention put $\lambda^{(0)}=1$.
First of all, note that $Z_{\alpha,r}(\rho||\sigma)^{\wedge k} = Z_{\alpha,r}(\rho^{\wedge k}||\sigma^{\wedge k})$.
Hence, we get
$$
\lambda_1(Z_{\alpha,r}(\rho||\sigma)^{\wedge k}) \approx \lambda_1(\rho^{\wedge k})^\alpha \;
\left(\left((\sigma^{\wedge k})^{-1/r}\right)_{1,1}\right)^{r(\alpha-1)}
$$
which means that
\be
\lambda^{(k)} :=\lambda_1\cdots \lambda_k(Z_\alpha(\rho||\sigma)) \approx
(\mu_1\cdots\mu_k)^\alpha \left(\det\left((\sigma^{-1/r})_{1:k,1:k}\right)\right)^{r(\alpha-1)}.\label{eq:lambdak}
\ee
A mathematically rigorous restatement of this approximate identity will be given below as the Approximation Lemma,
Lemma \ref{lem:app}.
For $k=d$, we actually obtain an exact expression as it reduces to the well-known statement that the determinant of
a product equals the product of the determinants:
$$
\lambda^{(d)} = \det(Z_\alpha(\rho||\sigma)) = (\det\rho)^\alpha \left(\det \sigma^{-1/r} \right)^{r(\alpha-1)}.
$$

It is now a simple matter to obtain an approximation for $\trace Z_{\alpha,r}(\rho||\sigma)$.
Indeed, by taking the quotients of successive $\lambda^{(k)}$ we get all the eigenvalues of
$Z_{\alpha,r}$: $\lambda^{(k)}/\lambda^{(k-1)} = \lambda_k(Z_{\alpha,r}(\rho||\sigma))$.
Summing these quotients then yields the trace of $Z_{\alpha,r}$:
$$
\trace Z_{\alpha,r}(\rho||\sigma) = \sum_{k=1}^d \lambda_k(Z_{\alpha,r}(\rho||\sigma))
= \lambda^{(1)} + \sum_{k=2}^d \frac{\lambda^{(k)}}{\lambda^{(k-1)}}.
$$
Inserting the approximation (\ref{eq:lambdak}) for $\lambda^{(k)}$ yields
\be
\trace Z_{\alpha,r}(\rho||\sigma) \approx \mu_1^\alpha \left((\sigma^{-1/r})_{1,1}\right)^{r(\alpha-1)}
+ \sum_{k=2}^d \mu_k^\alpha \left(\frac{ \det\left((\sigma^{-1/r})_{1:k,1:k}\right) }{ \det\left((\sigma^{-1/r})_{1:k-1,1:k-1}\right) }\right)^{r(\alpha-1)}.
\label{tracez}
\ee
Let us introduce the vector $\nu$ of leading principal minors of $\sigma^{-1/r}$ taken to the power $-r$, with
\be
\nu_k := \det\left((\sigma^{-1/r})_{1:k,1:k}\right)^{-r}.
\ee
Note that $\nu_d = \det \sigma$. In terms of these $\nu_k$, eq.\ (\ref{tracez})
can be rewritten more succinctly as
$$
\trace Z_{\alpha,r}(\rho||\sigma) \approx \mu_1^\alpha \nu_1^{1-\alpha}
+ \sum_{k=2}^d \mu_k^\alpha \left(\frac{ \nu_k }{ \nu_{k-1} }\right)^{1-\alpha}.
$$
One now recognizes the trace functional $f_{\alpha,z}$ in this formula,
between the state $\rho$ and a new positive definite matrix $\widehat\sigma$
that commutes with $\rho$ and that is given by
\be
\widehat\sigma = \diag_\rho(\nu_1, \nu_2/\nu_1, \nu_3/\nu_2,\ldots,\nu_d/\nu_{d-1}).
\ee
Here, $C=\diag_\rho(x_1,\ldots,x_d)$ denotes a matrix $C$ that is diagonal in the eigenbasis of $\rho$ and has diagonal elements $x_i$;
that is, $\trace P_i C=x_i$.

We then finally get, for $\alpha$ sufficiently close to 1:
\be
\trace Z_{\alpha,r}(\rho||\sigma) \approx \trace\rho^\alpha \widehat\sigma^{1-\alpha}.\label{eq:hattie}
\ee
The error in this approximation tends to $0$ exponentially fast as $\exp(-\kappa/|r(1-\alpha|))$, where $\kappa$ is a strictly positive constant
depending only on the eigenvalues $\mu_i$,
as shown in Lemma \ref{lem:app} below.
From (\ref{eq:hattie}) a closed form expression for the limit $\alpha\to 1$ of $D_{\alpha,r(\alpha-1)}$ can be found very easily,
and it simply gives the classical relative entropy between $\rho$ and $\widehat\sigma$.
We have therefore proven:
%%%
%%%
\begin{theorem}\label{th:D10}
Let $\rho$ be a positive semidefinite matrix with non-degenerate spectrum and let $\sigma$ be positive definite.
Let $r$ be a non-zero, finite real number.
Then
\bea
\lim_{\alpha\to 1} D_{\alpha,r(\alpha-1)}(\rho||\sigma) &=& D(\rho||\diag_\rho(\nu_1, \nu_2/\nu_1, \nu_3/\nu_2,\ldots,\nu_d/\nu_{d-1})),\nonumber\\
\mbox{ with } \nu_k &=& \det\left((\sigma^{-1/r})_{1:k,1:k}\right)^{-r},\quad k=1,\ldots,d.\label{eq:D10}
\eea
In particular, for $r=-1$,
\bea\label{eq18}
\lim_{\alpha\to 1} \widehat D_\alpha(\rho||\sigma) &=& D(\rho||\diag_\rho(\nu_1, \nu_2/\nu_1, \nu_3/\nu_2,\ldots,\nu_d/\nu_{d-1})),\nonumber\\
\mbox{ with }
\nu_k &=& \det\left(\sigma_{1:k,1:k}\right),\quad k=1,\ldots,d.
\eea
\end{theorem}
%%%
%%%
As a sanity check, we can consider what eq.~(\ref{eq18}) reduces to when $\rho$ and $\sigma$ commute.
In that case, $\sigma$ is diagonal in the eigenbasis of $\rho$, and its leading principal minors are
just the products of its $k$ first diagonal elements:
$\nu_k = \sigma_{1,1} \cdots \sigma_{k,k}$. Hence, the successive quotients $\nu_k/\nu_{k-1}$ reduce to $\sigma_{k,k}$,
and $\diag_\rho(\nu_1, \nu_2/\nu_1, \nu_3/\nu_2,\ldots,\nu_d/\nu_{d-1})$ simply turns into $\sigma$ itself.
We thus find that, in the commuting case, $\lim_{\alpha\to 1} D_{\alpha,r(\alpha-1)}(\rho||\sigma) = D(\rho||\sigma)$, as required.

\medskip

To complete the case of non-degenerate $\rho$, we now provide the Approximation Lemma in full detail.
\begin{lemma}[Approximation Lemma]\label{lem:app}
Let $A$ be a positive semidefinite matrix with its eigenvalues sorted in decreasing order denoted by $\mu_i$.
Let $B$ be a positive definite matrix and let $B_{1,1}$ be the upper left matrix element expressed in the eigenbasis of $A$.
%Let $0<\alpha<1$ and $\beta=\alpha/2(1-\alpha)$.
Let $\gamma>0$.
Then
$$
\lambda_1\left( (A^{\beta} BA^{\beta})^\gamma\right)
= \mu_1^{2\beta\gamma} (B_{1,1})^{\gamma} \left(1+c(\mu_2/\mu_1)^{2\beta}\right)^{\gamma},
$$
for some constant value of $c$ independent of $\gamma>0$.
\end{lemma}
In the proof of Theorem \ref{th:D10} we use the case $\gamma=r(1-\alpha)$, $\beta=\alpha/2\gamma$, $A=\rho$ and $B=\sigma^{-1/r}$.
The limit of $\alpha$ going to 1 (from below, if $r<0$, or from above, if $r>0$) corresponds to the limit $\gamma\to 0^+$,
and we get that for $\alpha$ tending to 1, $\lambda_1\left( (A^{\beta} BA^{\beta})^{1-\alpha}\right)$ tends to
$\mu_1^{2\beta\gamma} (B_{1,1})^{\gamma} = \mu_1^{\alpha} (B_{1,1})^{r(1-\alpha)}$
with an exponentially decreasing relative error $c\gamma\exp(-k/\gamma)$, with $k=|\log(\mu_2/\mu_1)|$, provided of course that
$\mu_1>\mu_2$, strictly.
This is because for $0\le x<1$ and very small $\gamma$ the function $(1+cx^{1/\gamma})^{\gamma}$ can be approximated as
\[
(1+cx^{1/\gamma})^{\gamma} \approx 1+c\gamma\exp(-|\log x|/\gamma).
\]

\textit{Proof.}
From the eigenvalue decomposition $A=\sum_{k=1}^d \mu_k P_k$ and the hypothesis $\mu_2<\mu_1$ we can write
$A=\mu_1 P_1 +X$ with $0\le X\le \mu_2(\id-P_1)$; note also that $X$ is orthogonal to $P_1$.
Thus,
\beas
\lambda_1\left(A^{\beta} BA^{\beta}\right)
&=& \lambda_1\left(B^{1/2} A^{2\beta} B^{1/2}\right) \\
&=& \lambda_1\left(B^{1/2} (\mu_1^{2\beta}P_1+X^{2\beta}) B^{1/2}\right).
\eeas
As the function that maps a Hermitian matrix to its largest eigenvalue is order-preserving and subadditive, this gives us
\bea
\lambda_1\left(A^{\beta} BA^{\beta}\right)
&\ge& \lambda_1\left(B^{1/2} \mu_1^{2\beta}P_1 B^{1/2}\right) \nonumber \\
&=& \mu_1^{2\beta} B_{1,1} \label{eq:app1}
\eea
and
\bea
\lambda_1\left(A^{\beta} BA^{\beta}\right)
&\le& \lambda_1\left(B^{1/2} (\mu_1^{2\beta}P_1+\mu_2^{2\beta}(\id-P_1)) B^{1/2}\right) \nonumber\\
&\le& \mu_1^{2\beta} \lambda_1\left(B^{1/2} P_1 B^{1/2}\right)
      +\mu_2^{2\beta}\lambda_1\left(B^{1/2} (\id-P_1) B^{1/2}\right) \nonumber\\
&\le& \mu_1^{2\beta} B_{1,1} + \mu_2^{2\beta}\lambda_1(B) \nonumber\\
&=& \mu_1^{2\beta} B_{1,1}\left(1+\frac{\lambda_1(B)}{B_{1,1}}(\mu_2/\mu_1)^{2\beta}\right).\label{eq:app2}
\eea
Since $B>0$, we have $B_{1,1}>0$ and the division can be done.
Bracketing inequalities (\ref{eq:app1}) and (\ref{eq:app2}) can be combined as a single equality
by introducing a constant $c$ such that
$$
\lambda_1\left(A^{\beta} BA^{\beta}\right) = \mu_1^{2\beta} B_{1,1}\left(1+c(\mu_2/\mu_1)^{2\beta}\right),
$$
and imposing that $c$ lies between $0$ and $\lambda_1(B)/B_{1,1}$.

Raising all expressions to the (positive) power $\gamma$ yields the equality of the lemma
$$
\lambda_1\left( (A^{\beta} BA^{\beta})^{\gamma}\right)
= \mu_1^{2\gamma\beta} (B_{1,1})^{\gamma} \left(1+c (\mu_2/\mu_1)^{2\beta}\right)^{\gamma}.
$$
\qed

\bigskip

Let us now consider what happens when the spectrum of $\rho$ is degenerate, and whether $D_{\alpha,r(\alpha-1)}(\rho||\sigma)$ is continuous
in $\rho$ and $\sigma$ (with $\supp\rho\le\supp\sigma$) in the limit $\alpha\to 1$.
It is clear from the definition that it is continuous for all $\alpha\neq1$.
Thus, if we can show that (\ref{eq:D10}) has a continuous extension, one that includes degenerate $\rho$ as well, then $D_{\alpha,r(\alpha-1)}$
is indeed continuous in the limit $\alpha\to 1$.

Let us therefore consider (\ref{eq:D10}) at face value (without looking back at the arguments that were used to derive it)
and see whether it is even well-defined for degenerate $\rho$.
This is not immediately clear because of the formula's non-trivial dependence on the eigenbasis of $\rho$: when the spectrum of $\rho$ is
degenerate, $\rho$ has an infinity of allowed eigenbases, and the question arises whether the choice of basis affects the outcome.
It turns out, however, that it does not, as the eigenvalue multiplicity `both gives and takes', as explained below.

For the sake of concreteness, let us take a $\rho$ for which $\mu_1$ has multiplicity 2. Then $P_1$ is a 2-dimensional projector, and
any pair of orthonormal vectors in the corresponding subspace can serve as basis elements.
For every such basis, one gets a different matrix representation of $\sigma$.
This can be recast as fixing one such representation of $\sigma$ and letting a $2\times 2$ unitary
matrix $U$ act on its upper left $2\times 2$ block.
Consequently, $\nu_1$ depends on $U$ whereas the other $\nu_i$ are independent from $U$, due to unitary invariance of the determinant.
However, whereas this clearly affects the first two elements in the resulting
$$
\widehat\sigma = \diag_\rho(\nu_1, \nu_2/\nu_1, \nu_3/\nu_2,\ldots,\nu_d/\nu_{d-1}),
$$
this is actually compensated for by the multiplicity of $\mu_1$.
The first two terms in the formula for $D(\rho||\widehat\sigma)$ are
$$
D(\rho||\widehat\sigma) = \mu_1(\log\mu_1-\log\nu_1) + \mu_1(\log\mu_1-\log(\nu_2/\nu_1)) + \cdots
$$
and this simplifies to
$$
D(\rho||\widehat\sigma) = 2\mu_1\log\mu_1-\mu_1\log\nu_2 + \cdots
$$
which is independent of $\nu_1$.

One checks that this argument generalizes to all possible multiplicities.
In fact, an equivalent formula for $D(\rho||\widehat\sigma)$ is
\be
D(\rho||\widehat\sigma) = -S(\rho)-\mu_d\log\det\sigma - \sum_{i=1}^{d-1}(\mu_i-\mu_{i+1})\log\nu_i,\label{eq:newhat}
\ee
where $S(\rho)= - \tr \rho \log \rho$ is the von Neumann entropy of $\rho$.
%Additionally, in terms of the spectral representation of $\rho$, where the \emph{distinct}, sorted eigenvalues of $\rho$ are denoted by $\lambda_i$
%and the projectors on the eigenspaces of $\rho$ by $P_i$,
%\be
%\rho = \sum_{i=1}^{r} \lambda_i P_i^* P_i,
%\ee
%we have
%\be
%D(\rho||\widehat\sigma) = -S(\rho)-\lambda_r\log\det\sigma - \sum_{i=1}^{r-1}(\lambda_i-\lambda_{i+1})\log\det(Q_i\sigma Q_i^*),
%\mbox{ with } Q_i:=\sum_{k=1}^i P_k.
%\label{eq:newhat2}
%\ee
The upshot is that $D(\rho||\widehat\sigma)$ is independent of those elements $\nu_i$
that are dependent on a freedom of choice of basis caused by degeneracy
of $\mu_i$. % and \emph{only} depend on the eigenprojections $P_i$.
This implies that $D(\rho||\widehat\sigma)$ is continuous in $\rho$ and $\sigma$ since every term in (\ref{eq:newhat}) is continuous, as we now show.
Indeed, the von Neumann entropy is well-known to be continuous (in the sense of Fannes), and $\mu_d$ and $\nu_d=\det\sigma$ are continuous as well since
eigenvalues of a matrix depend continuously on the entries of a matrix (\cite{HJI}, Appendix D).
The only potential problems stem from the terms $(\mu_i-\mu_{i+1})\log\nu_i$ as they explicitly depend on the eigenprojections of $\rho$.

To see the problem, consider the example of a positive semidefinite matrix $\rho$ parameterized by the variable $x$,
$\rho(x)=\diag(1+x,1-x)$, with $0<|x|<1$. Then for $x>0$, $P_1 = \diag(1,0)$ whereas for $x<0$, $P_1=\diag(0,1)$. Thus
for almost all $\sigma$, $\nu_1(x)$ has a discontinuity at $x=0$.
However, these discontinuities only occur at the so-called \emph{exceptional points} of $\rho(x)$, the points where some eigenvalues coincide,
a.k.a.\ level-crossings in physics terminology. This is because eigenprojections of Hermitian $\rho(x)$
are holomorphic functions of $x$ (\cite{kato}, Chapter II,
Theorem 6.1). The discontinuities occur because the ordering of the eigenvalues changes
at a level-crossing, and the eigenprojections get swapped accordingly,
as in the example. The terms $(\mu_i-\mu_{i+1})\log\nu_i$, however, remain continuous,
since any level-crossing affecting $\nu_i$ occurs when the prefactor $\mu_i-\mu_{i+1}$ becomes zero, which cancels the
discontinuity in $\nu_i$ (while still leaving a discontinuity in the derivative).

We have thus finally proven:
\begin{theorem}\label{thm_3}
The statement from Theorem \ref{th:D10} still holds when the spectrum of $\rho$ is degenerate, in the sense that (\ref{eq:D10}) has to be
interpreted as (\ref{eq:newhat}).
The limit $\lim_{\alpha\to 1} D_{\alpha,r(\alpha-1)}(\rho||\sigma)$ exists as a continuous (but not necessarily smooth) function of $\rho$ and $\sigma$.
\end{theorem}

%%%%%%%%%%%%%%%%%%%%%%%%%%%%%%%%%%%%%%%%%%%%%%%%%%%%%%%%%%%%%%%%%%%%%%%%%%%%%%%%%%%%%%%%%%%%%%%%%%%%%%%%%%%%%%%%%%%%%%%%%%%%%%%%%%%%%%%%%%%%%%
%%%%%%%%%%%%%%%%%%%%%%%%%%%%%%%%%%%%%%%%%%%%%%%%%%%%%%%%%%%%%%%%%%%%%%%%%%%%%%%%%%%%%%%%%%%%%%%%%%%%%%%%%%%%%%%%%%%%%%%%%%%%%%%%%%%%%%%%%%%%%%
\section{The case of infinite $z$}\label{sec:limitinf}
In this section we study the behaviour of $D_{\alpha,z}$ for $z$ going to infinity.
As in the previous section we first consider the parametrization $z=r(\alpha-1)$, with $r>0$, and take the limit of
$D_{\alpha,r(\alpha-1)}$ as $\alpha$ tends to $+\infty$.

Noting that the operator norm is the limit of the Schatten $q$-norm as $q$ tends to $+\infty$, we obtain from \reff{eq:frs1},
\beas
\lim_{\alpha\to +\infty} D_{\alpha,r(\alpha-1)}(\rho||\sigma)
&=& \lim_{\alpha\to +\infty} \frac{1}{\alpha-1} \log\trace(\rho^{\alpha/r(\alpha-1)}\sigma^{-1/r})^{r(\alpha-1)} \\
&=& \lim_{\alpha\to +\infty} \log||(\rho^{\alpha/2r(\alpha-1)} \sigma^{-1/r} \rho^{\alpha/2r(\alpha-1)})^r||_{\alpha-1} \\
&=& \log||(\rho^{1/2r} \sigma^{-1/r} \rho^{1/2r})^r||_{\infty} \\
&=& \log||\rho^{1/2r} \sigma^{-1/r} \rho^{1/2r}||_{\infty}^r \\
&=& r \log||\rho^{1/2r} \sigma^{-1/r} \rho^{1/2r}||_{\infty}.
\eeas
Now the operator norm of a positive semidefinite matrix $X$ equals the largest eigenvalue of $X$, which in turn is the smallest value
of $\lambda$ such that $X\le \lambda I$.
In the present case, this condition is $\rho^{1/2r} \sigma^{-1/r} \rho^{1/2r}\le\lambda I$, which is equivalent to $\lambda\sigma^{1/r}\ge\rho^{1/r}$. Hence,
$$\log||\rho^{1/2r} \sigma^{-1/r} \rho^{1/2r}||_{\infty}
= \log \min_\lambda\{\lambda: \lambda\sigma^{1/r}\ge\rho^{1/r}\} = D_{\max}(\rho^{1/r}||\sigma^{1/r}).$$
% Hence, in the limit, $D_{\alpha,2r(\alpha-1)}$ tends to the max-relative entropy, and
Thus we arrive at the following theorem:
\begin{theorem}\label{thm_4}
Let $\rho$ be a positive semidefinite matrix and let $\sigma$ be positive definite.
Then for a non-zero, finite real number $r$,
\be
\lim_{\alpha\to +\infty} D_{\alpha,r(\alpha-1)}(\rho||\sigma)
= r D_{\max}(\rho^{1/r}||\sigma^{1/r}).
\ee
In particular, for $r=1$
$$\lim_{\alpha\to +\infty} \widetilde D_\alpha(\rho||\sigma) = D_{\max}(\rho||\sigma).$$
\end{theorem}
\medskip

For $\alpha\to -\infty$, which necessitates the stronger restriction on the supports $\supp\rho=\supp\sigma$, a similar treatment yields the result that
for $r<0$,
\be
\lim_{\alpha\to -\infty} D_{\alpha,r(\alpha-1)}(\rho||\sigma) = r D_{\max}(\sigma^{-1/r} || \rho^{-1/r})
\ee
and, for $r=-1$,
\be
\lim_{\alpha\to -\infty} \widehat D_\alpha(\rho||\sigma) = - D_{\max}(\sigma || \rho).
\ee

Finally, we study the limit $z\to\infty$ when $\alpha$ is kept fixed (and finite).
Let us first consider the case where $\supp\rho=\supp\sigma$.
Using the well-known Lie-Trotter product formula (see, e.g.\ \cite{bhatia}, Theorem IX.1.3),
according to which $\lim_{m\to\infty}(\exp(A/m)\exp(B/m))^m=\exp(A+B)$
for any two matrices $A$ and $B$, we easily obtain (with $A=\log\rho^\alpha$ and $B=\log\sigma^{1-\alpha}$), for $\alpha\neq1$,
\be
\lim_{z\to\infty} D_{\alpha,z}(\rho||\sigma) = \frac{1}{\alpha-1}\log\trace\exp(\alpha\log\rho+(1-\alpha)\log\sigma).
\ee
In the limit $\alpha\to1$, we use l'H\^opital's rule and the fact that $(d/d\alpha) \trace\exp(X+\alpha Y)=\trace Y\exp(X+\alpha Y)$
to obtain
\be
\lim_{\alpha\to1} \lim_{z\to\infty} D_{\alpha,z}(\rho||\sigma) = D(\rho||\sigma).
\ee
When $\supp\rho$ is a proper subset of $\supp\sigma$, the same formulas hold except for the fact that
we have to restrict $\log\sigma$ to $\supp\rho$ (more generally, both $\log\rho$ and $\log\sigma$ have to be restricted to the intersection of the
supports of $\rho$ and $\sigma$). This was proven by Hiai and Petz in \cite{hiaipetz}.

After the first draft of the present paper had been circulated, Lin and Tomamichel have shown \cite{lintoma}
that the relative entropy is recovered more generally when
$\alpha$ goes to 1 and $z$ is taken to be $z=g(\alpha)$, for any continuously differentiable function $g$ such that $g(1)\neq 0$.
%%%%%%%%%%%%%%%%%%%%%%%%%%%%%%%%%%%%%%%%%%%%%%%%%%%%%%%%%%%%%%%%%%%%%%%%%%%%%%%%%%%%%%%%%%%%%%%%%%%%%%%%%%%%%%%%%%%%%%%%%%%%%%%%%%%%%%%%%%%%%%
\section{Limiting case $z=\alpha\to0^+$}\label{sec:lim00}
In this section, we answer the question: what is the limit of $\widetilde D_\alpha$ as $\alpha$ tends to 0; that is, what is
$$
\lim_{\alpha\to 0}D_{\alpha,\alpha}(\rho||\sigma) = -\log \lim_{\alpha\to 0}f_{\alpha,\alpha}(\rho||\sigma)?
$$
As always, we assume that $\sigma$ is full rank. We will also assume first that the spectrum of $\sigma$ is non-degenerate.

The answer to this question is easy when $\rho$ and $\sigma$ commute.
Choosing a basis in which both states are diagonal, with diagonal elements given by $\rho_i$ and $\sigma_i$, respectively, the limit is given by
\beas
\lim_{\alpha\to 0}f_{\alpha,\alpha}(\rho||\sigma) &=& \lim_{\alpha\to 0} \sum_{i=1}^d \rho_i^\alpha\sigma_i^{1-\alpha} \\
&=& \sum_{i}\sigma_i: \rho_i\neq0.
\eeas
In terms of the projector on the support of $\rho$, which we denote by $\Pi_\rho$, we write this as
$$
\lim_{\alpha\to 0}f_{\alpha,\alpha}(\rho||\sigma) = \trace \Pi_\rho \sigma.
$$

\bigskip

To answer the question in the general case, we will first show that the answer does not depend on $\rho$ itself, but only on $\Pi_\rho$, and of course
also on $\sigma$.
To do so, we consider the particular expression
$$
\lim_{\alpha\to 0} f_{\alpha,\alpha}(\rho||\sigma)
= \lim_{\alpha\to 0} \trace(\sigma^{1/2\alpha} \rho \sigma^{1/2\alpha})^\alpha.
$$
Let $\mu$ be the smallest non-zero eigenvalue of $\rho$.
Then we have the inclusion $\mu\Pi_\rho \le \rho \le \Pi_\rho$.
This implies
$$
\mu^\alpha \; \trace(\sigma^{1/2\alpha} \Pi_\rho \sigma^{1/2\alpha})^\alpha
\le
\trace(\sigma^{1/2\alpha} \rho \sigma^{1/2\alpha})^\alpha
\le \trace(\sigma^{1/2\alpha} \Pi_\rho \sigma^{1/2\alpha})^\alpha.
$$
In the limit of $\alpha\to0$, $\mu^\alpha$ of course tends to 1, so that both sides of the inclusion become equal
and we have the identity
$$
\lim_{\alpha\to 0} \trace(\sigma^{1/2\alpha} \rho \sigma^{1/2\alpha})^\alpha
= \lim_{\alpha\to 0} \trace(\sigma^{1/2\alpha} \Pi_\rho \sigma^{1/2\alpha})^\alpha.
$$

\bigskip

For the remainder of the argument, we will work in a basis in which $\Pi_\rho$ is diagonal, and given by $\id_r\oplus 0$, where $r$ is the rank of $\rho$.
Furthermore, we switch from one representation of $f_{\alpha,\alpha}$ to another, namely
$$
\lim_{\alpha\to 0} f_{\alpha,\alpha}(\rho||\sigma) =
\lim_{\alpha\to 0} \trace(\Pi_\rho \sigma^{1/\alpha}\Pi_\rho)^\alpha.
$$
We will also employ the spectral decomposition of $\sigma$, which we consider to be given by
$$
\sigma = U\Lambda U^* =\sum_{i=1}^d \lambda_i |u_i\rangle \langle u_i|,
$$
where the eigenvalues are sorted in descending order as $\lambda_1 > \lambda_2 >\cdots >\lambda_d$.
To deal with the expression $\Pi_\rho \sigma^{1/\alpha}\Pi_\rho$, we will finally define the restriction of the eigenvectors to the support of $\rho$:
$$
|u_i\rangle \mapsto |\widetilde u_i\rangle := \Pi_\rho |u_i\rangle.
$$
With this definition, we have
$$
\Pi_\rho \sigma^{1/\alpha}\Pi_\rho = \sum_{i=1}^d \lambda_i^{1/\alpha} |\widetilde u_i\rangle \langle \widetilde u_i|.
$$
It goes without saying that the vectors $|\widetilde u_i\rangle$ in general no longer form an orthonormal set,
and the quantities $\lambda_i^{1/\alpha}$
are not eigenvalues of $\Pi_\rho \sigma^{1/\alpha}\Pi_\rho$.

Let us first try and find an expression for the largest eigenvalue $\mu_1$ of $Z_\alpha:=(\Pi_\rho \sigma^{1/\alpha}\Pi_\rho)^\alpha$ in the limit $\alpha\to 0^+$.
Given that the spectrum of $\sigma$ is non-degenerate, the main contribution to $\Pi_\rho \sigma^{1/\alpha}\Pi_\rho$ as $\alpha\to0^+$
will come from $\lambda_1$, and is given by $\lambda_1^{1/\alpha} |\widetilde u_1\rangle \langle \widetilde u_1|$.
That is true, of course, only if $|\widetilde u_1\rangle$ is not the zero vector ($|\widetilde u_1\rangle=0$ if $|u_1\rangle$
lies outside the support of $\rho$).
We therefore have to correct our statement and say:
the main contribution to $\Pi_\rho \sigma^{1/\alpha}\Pi_\rho$
will come from $\lambda_{i_1}$, and is given by $\lambda_{i_1}^{1/\alpha} |\widetilde u_{i_1}\rangle \langle \widetilde u_{i_1}|$,
where $i_1$ is the first index value for which $|\widetilde u_i\rangle\neq0$.
The limit can now be calculated easily, and we get
$$
\mu_1 = \lim_{\alpha\to 0} \lambda_{i_1} ||\;|\widetilde u_{i_1}\rangle \langle \widetilde u_{i_1}|\;||^\alpha
= \lambda_{i_1} = \max_{i_1}\lambda_{i_1}: |\widetilde u_{i_1}\rangle\neq0
$$

Next, we calculate the product of the two largest eigenvalues of $Z_\alpha$, $\mu_1\mu_2$, in the limit $\alpha\to 0^+$.
Using the Weyl-trick, this reduces to the largest eigenvalue of the second antisymmetric tensor power, and using the formula just obtained
we find
$$
\mu_1\mu_2 = \max_{i_1,i_2}\lambda_{i_1}\lambda_{i_2}: |\widetilde u_{i_1}\rangle\wedge |\widetilde u_{i_2}\rangle\neq0.
$$
The latter condition amounts to the two vectors $|\widetilde u_{i_1}\rangle$ and $|\widetilde u_{i_2}\rangle$ being linearly independent.
For $\mu_1\mu_2\mu_3$ we similarly obtain
$$
\mu_1\mu_2\mu_3 = \max_{i_1,i_2,i_3}\lambda_{i_1}\lambda_{i_2}\lambda_{i_3}: |\widetilde u_{i_1}\rangle, |\widetilde u_{i_2}\rangle,
|\widetilde u_{i_3}\rangle \mbox{ linearly independent},
$$
and so on either until $\mu_1\mu_2\cdots\mu_r$ has been obtained, or no further linearly independent vectors can be added to the set.
That is, the process stops at $\mu_1\mu_2\cdots \mu_s$, where $s$ is the rank of $\Pi_\rho \sigma$ (clearly, $s\le r$).

By successive divisions we then find the separate $\mu_i$, for $i=1,2,\ldots,s$.
What we are after is the sum of these $\mu_i$, and this sum is simply given by
$$
\sum_{i=1}^s \mu_i =
\max_{i_1,i_2,\ldots,i_s}
\sum_{j=1}^s \lambda_{i_j}: \{|\widetilde u_{i_j}\rangle\}\mbox{ linearly independent}.
$$
A convenient way to find these linearly independent vectors is to use Gaussian elimination, under the guise of the
Row-Echelon normal Form (REF) procedure (well-known from any introductory Linear Algebra course).
The indices $i_j$ of the formula are the column indices of those columns that contain a row-leading entry (that is, the first non-zero entry in some row)
in the row-echelon normal form of the matrix $\Pi_\rho U$.

We have therefore proven:
\be
\lim_{\alpha\to 0}f_{\alpha,\alpha}(\rho||\sigma) = \sum_{j=1}^s \lambda_{i_j},
\ee
where the $\lambda_i$ are the eigenvalues of $\sigma$, and the indices $i_j$ can be found from the following procedure:
calculate the row-echelon form $R$ of the matrix $\Pi_\rho U$ (expressed in an eigenbasis of $\rho$).
For every row of $R$, determine at which column the first non-zero entry appears; these column indices are the sought values of $i_j$
and $s$ is the number of non-zero rows in $R$.

\bigskip

The result just obtained still holds in the case when the spectrum of $\sigma$ is degenerate.
Suppose a certain eigenvalue of $\sigma$ has multiplicity $k$.
Let $S$ be the subspace that is the projection of this $k$-dimensional eigenspace to the support of $\rho$.
The problem is that one can choose among an infinite number of bases for $S$;
which basis contains the highest number of vectors that are independent from the $u_{i_j}$ that we already had?
The answer is simple: that number is really basis independent and only depends on the dimension of the intersection of $S$ with the subspace $P$ spanned
by these $u_{i_j}$.
Thus any basis should do, and the formula remains as it stands.

\bigskip

We finish this section with a simple example of the procedure just described.
Let $\rho$ and $\sigma$ be $4$-dimensional states where $\sigma$ is full rank and has non-degenerate spectrum, and $\rho$ has rank 2.
In terms of the eigenbasis of $\rho$, the projector $\Pi_\rho$ is represented by the diagonal matrix $\Pi_\rho=\diag(1,1,0,0)$.
Furthermore, let $\sigma$ have spectral decomposition $\sigma=\sum_{i=1}^4 \lambda_i |u_i\rangle\langle u_i|$ where the eigenvectors $|u_i\rangle$ are the
columns of the unitary matrix
$$
U = \half\left(
\begin{array}{rrrr}
1& 1&  1& 1 \\
1& 1& -1& -1\\
1& -1&  1& -1\\
1& -1& -1& 1
\end{array}
\right).
$$
Thus, the matrix $\Pi_\rho U$ (after deleting the rows that are completely zero) and its REF are given by
$$
\Pi_\rho U = \half\left(
\begin{array}{rrrr}
1& 1&  1& 1 \\
1& 1& -1& -1
\end{array}
\right)
\mbox{ and }
\mbox{REF}(\Pi_\rho U) = \half\left(
\begin{array}{rrrr}
1& 1&  1& 1 \\
0& 0& -2& -2
\end{array}
\right).
$$
The row-leader of row 1 is in column 1, and the one of row 2 is in column 3. Therefore, we put $i_1=1$ and $i_2=3$,
so that the value of $\lim_{\alpha\to 0}f_{\alpha,\alpha}(\rho||\sigma) = \sum_{i=1}^s \mu_i$ is given by $\lambda_1+\lambda_3$.
%%%%%%%%%%%%%%%%%%%%%%%%%%%%%%%%%%%%%%%%%%%%%%%%%%%%%%%%%%%%%%%%%%%%%%%%%%%%%%%%%%%%%%%%%%%%%%%%%%%%%%%%%%%%%%%%%%%%%%%%%%%%%%%%%%%%%%%%%%%%%%
%%%%%%%%%%%%%%%%%%%%%%%%%%%%%%%%%%%%%%%%%%%%%%%%%%%%%%%%%%%%%%%%%%%%%%%%%%%%%%%%%%%%%%%%%%%%%%%%%%%%%%%%%%%%%%%%%%%%%%%%%%%%%%%%%%%%%%%%%%%%%%
\section{Proof of concavity of the trace functional $f_{\alpha,z}$}\label{proof_epstein}
For convenience, we will consider the trace functional $f_{\alpha,z}$ in its re-parameterized form
\be
f_{p,q}(A;K) := \trace(A^p K A^q K^*)^{1/(p+q)},
\ee
where $p=\alpha/z$ and $q=(1-\alpha)/z$;
thus $z=1/(p+q)$ and $\alpha=p/(p+q)$.
To lighten the notations we will henceforth just write $f_{p,q}(A)$ as we always keep $K$ fixed.

In this section we shall provide a detailed proof of the following:
\begin{theorem}[Hiai]\label{th:concave}
The trace functional $f_{p,q}(A)$ is concave on the set of positive semidefinite $d\times d$ matrices for $0<p,q\le 1$.
\end{theorem}
This theorem is a special case of Theorem 1.1 in \cite{hiai13} (namely the case $s=1/(p+q)$ and replacing the positive linear maps $\Phi$ and $\Psi$ by the
identity map),
but as the proof of that Theorem is rather complicated, in part due to its generality, we provide a more direct proof of our own.
Furthermore, the proof technique is based on Epstein's proof \cite{epstein} of Lieb's concavity theorem, and
as the exposition in \cite{epstein} is rather terse
we also felt the need to expand on the details where necessary and simplify the proof where possible.
We hope that our exposition will make this powerful technique more widely accessible.
\subsection{Reduction step}
The function $f_{p,q}$ is concave over the set of positive semidefinite matrices if and only if
for all $A_1,A_2\ge0$ and for all $s\in[0,1]$
\[
f_{p,q}(s A_1+(1-s)A_2) \ge s f_{p,q}(A_1) +(1-s) f_{p,q}(A_2).
\]
Since $f_{p,q}$ is homogeneous, this can be rewritten as
\[
f_{p,q}(A_1+t A_2) \ge f_{p,q}(A_1) + t f_{p,q}(A_2),
\]
where $t=(1-s)/s$, hence this statement has to hold for all $t\ge0$.
Introducing the function
\[
F(t) = f_{p,q}(A_1+t A_2),
\]
where $A_1$ and $A_2$ are kept fixed, we have to show that
\[
F(t) \ge \alpha+\beta t,
\]
where
\beas
\alpha &=& f_{p,q}(A_1) = F(0) \\
\beta  &=& f_{p,q}(A_2) = \lim_{x\to +\infty} \frac{F(x)}{x}.
\eeas

The main idea behind Epstein's technique is to show that $F(t)$ is a so-called Pick function.
A Pick function (a.k.a.\ Nevanlinna function or Herglotz function) is a function of a complex variable that is
holomorphic in the upper half-plane and maps it into itself, i.e.\ for all $z$ with $\im z>0$, $F(z)$ has non-negative imaginary part.
Such functions can be continued analytically by reflection (that is, $F(\overline z)=\overline{F(z)}$)
to the lower half-plane across a certain real interval.
For real $a<b$ one defines the subclass $P(a,b)$ of Pick functions for which this continuation is across the interval $(a,b)$.
For more details about Pick functions, and their relation to operator monotone functions,
we refer to \cite{donoghue}, especially its Chapters II and IX.

In particular, we will show that $F$ is in $P(0,+\infty)$ (or even in $P(\tau,+\infty)$, where $\tau$ is a negative number depending on $A_1$ and $A_2$).
Functions in this class have the integral representation
\be
F(t) = a+b t+\int_{-\infty}^0 \left(\frac{1}{x-t}-\frac{x}{x^2+1}\right) d\mu(x) \label{eq:Frep}
\ee
where $a$ is real, $b\ge0$ and $d\mu(x)$ is a positive measure on the interval $(-\infty,0]$ satisfying the condition
$\int_{-\infty}^0 (1+x^2)^{-1}\;d\mu(x)<+\infty$.
Then
\[
F(t)-F(0) = b t +\int_{-\infty}^0 \left(\frac{1}{x-t}-\frac{1}{x}\right) d\mu(x).
\]
Clearly, the integrand is positive for all $t>0$, and
\[
\beta = \lim_{t\to+\infty}(F(t)-F(0))/t = b.
\]
Thus, $F$ satisfies the inequality $F(t)\ge F(0)+ \beta t$, as required.

The remainder of this section will be devoted to proving that $F$ is indeed in $P(0,+\infty)$.
To this purpose, a number of technical tools first have to be introduced.

\textbf{Remark.}
As (\ref{eq:Frep}) is the canonical representation of a function that is operator monotone on the interval $[0,+\infty)$
it is not just concave, but also operator concave. %%% for what it is worth
More precisely, the extension of the function $F(t)=f_{p,q}(A_1+tA_2)$ to matrix arguments, namely
the function $Z\mapsto F(Z):=f_{p,q}(\id\otimes A_1+Z\otimes A_2;\id\otimes K)$, is operator concave (and operator monotone)
over the positive semidefinite matrices, for any fixed $A_1,A_2\ge 0$, and with $K$ replaced by $\id\otimes K$.

%%%%%%
\subsection{Technical Tools}
For any complex number $z$, and a complex matrix $C$ in $M_d(\mathbb{C})$ we use the notation $(C + z)$ to denote the matrix $(C+zI)$,
where $I$ denotes the identity matrix. We denote the conjugate transpose of a matrix $C$ by $C^*$ and use $C^{-*}$ as a shorthand
for $(C^{-1})^* = (C^*)^{-1}$. Furthermore, we denote the spectrum of any complex matrix $C \in M_d(\mathbb{C})$ as ${\rm{Sp}}\, C$.

%\subsubsection{Holomorphic matrix-valued functions}
The definition of holomorphy for complex-valued functions extends to
vector-valued and operator-valued functions in a straightforward way (see for example, Chapter 1 in \cite{KR}).
In particular, a matrix-valued function $C(z)$ is holomorphic over $D$ if and only if it is elementwise holomorphic; that is, if, for all $i$ and $j$,
$C_{ij}(z)$ is holomorphic over an open subset containing $D$.

If $C(z)$ and $D(z)$ are two holomorphic matrix-valued functions, then so is their matrix product $C(z)D(z)$,
and the inverse $C(z)^{-1}$ is holomorphic except where $C(z)$ is not invertible.

\subsubsection{Real and imaginary part of a matrix}
The real and imaginary part of a general complex matrix $C$ are defined as
\be
\re C := \frac{C+C^*}{2},\quad \im C:= \frac{C-C^*}{2i},
\ee
which are both Hermitian matrices.
Conversely, any matrix $C$ can be written as $C=\re C + i \im C$, a decomposition known as the Cartesian decomposition.

We summarize a number of properties that will be needed.

\begin{lemma}\label{lem:conjug}
For any pair of matrices $C$ and $K$,
$$
\re(KCK^*) = K(\re C) K^*,\mbox{ and }\im(KCK^*) = K(\im C) K^*.
$$
\end{lemma}
\textit{Proof.}
By the definition of the real part,
$$
\re(KCK^*) = \frac{KCK^* + (KCK^*)^*}{2} = \frac{KCK^* + KC^*K^*}{2} = K\frac{C+C^*}{2}K^* = K(\re C) K^*.
$$
The proof for the imaginary part is completely similar.
\qed

\begin{lemma}\label{lem:inv}
If $\im C>0$, then $C$ is invertible.
If $\re C>0$, then $C$ is invertible.
\end{lemma}
\textit{Proof.}
Let $A=\re C$ and $B=\im C$, so that $C=A+iB$.
By the hypothesis, $B>0$. Hence, $B$ is invertible and its positive square root exists too.
We can therefore write $C$ as
$$
C = B^{1/2}(B^{-1/2} A B^{-1/2} + i\id)B^{1/2}.
$$
As $B^{-1/2} A B^{-1/2}$ is Hermitian, it has real eigenvalues.
Any matrix commutes with the identity matrix. Thus, the eigenvalues of $B^{-1/2} A B^{-1/2} + i\id$
are of the form $\lambda+i$ with $\lambda\in \R$.
No eigenvalue can therefore be zero, so that $B^{-1/2} A B^{-1/2} + i$ is an invertible matrix.
As $B^{1/2}$ is invertible too, this shows that
$C$ is indeed invertible.

The statement for $\re C$ follows from this by noting that $\re C = \im(i C)$.
\qed

\begin{lemma}\label{lem:invert}
For any invertible matrix $C$,
\bea
\re C^{-1} &=& C^{-1} (\re C) C^{-*} \\
\im C^{-1} &=& -C^{-1} (\im C) C^{-*}.
\eea
\end{lemma}
\textit{Proof.}
If $C$ is invertible, so is $C^*$ and we can write $C^{-1}$ as
$$
C^{-1} = C^{-1} C^* C^{-*}.
$$
Using Lemma \ref{lem:conjug} with $K=C^{-1}$, we get
$$
\re C^{-1} = \re(C^{-1} C^* C^{-*}) = C^{-1} \re(C^*) C^{-*} = C^{-1} \re C C^{-*}
$$
and
$$
\im C^{-1} =  C^{-1} \im(C^*) C^{-*} = - C^{-1} \im C C^{-*}.
$$
\qed

\begin{lemma}\label{lem:spAB}
Let $A,B$ have positive real part. Then any real eigenvalue of $AB$ must be positive.
\end{lemma}
\textit{Proof.}
By Lemmas \ref{lem:inv} and \ref{lem:invert}, $A$ and $B$ are invertible and $\re B^{-1}>0$.
Let $x\ge0$. Then $\re (A+xB^{-1}) = \re A+x\re B^{-1}>0$, so that $A+x B^{-1}$
is invertible (again by Lemma \ref{lem:inv}). Hence, $AB+x$ is invertible too, and $-x$ can not be an eigenvalue of $AB$.
It follows that no eigenvalue of $AB$ lies on $\R^-$.
\qed

%%%%%%
\subsubsection{Complex segments}
Following \cite{epstein} we denote by $\cI^+(\C)$ ($\C^+$ in Hiai's notation) the open half-plane of complex numbers with positive imaginary part:
\be
\cI^+(\C) := \{z:\im z>0\}.
\ee
One can also define the open half-planes $\cI^-$ and $\cR^+$ as the sets of complex numbers
with negative imaginary part and positive real part, respectively.

These definitions generalize to complex matrices:
\be
\cI^+(\M_d(\C)) := \{C\in \M_d(\C):\im C>0\}.
\ee
We shall drop the arguments and write $\cI^+$ if it is clear from the context which set is meant.

Next, we introduce a new notation of our own, not present in \cite{epstein}.
Given two angles $\alpha$ and $\beta$, with $-\pi\le\alpha<\beta\le\pi$ and $\beta-\alpha\le\pi$, we denote the open segment
of the cut plane consisting of non-zero complex numbers whose complex argument is (strictly) between $\alpha$ and $\beta$ as
\be
\cS_{\alpha,\beta}(\C) := \{z=r e^{i\theta}: r>0, \alpha<\theta<\beta\}.
\ee
When $\beta-\alpha=\pi$, the segment is an open half-plane and can also be defined in terms of $\cI^+$ as
$$
\cS_{\alpha,\alpha+\pi}(\C) = \{z: e^{-i\alpha}z\in \cI^+\}.
$$
In particular, note that $\cI^+ = \cS_{0,\pi}$, $\cI^- = \cS_{-\pi,0}$ and $\cR^+=\cS_{-\pi/2,\pi/2}$.

Complex segments can equivalently be defined as the intersection of two open half-planes:
\bea
\cS_{\alpha,\beta}
&=& \cS_{\alpha,\alpha+\pi}(\C) \cap \cS_{\beta-\pi,\beta}(\C) \\
&=& \{z: e^{-i\alpha}z\in \cI^+ \} \cap \{z: e^{-i\beta}z\in \cI^- \} \nonumber \\
&=& \{z: e^{-i\alpha}z\in \cI^+ \mbox{ and } e^{-i\beta}z\in \cI^- \}.\label{eq:acutescalar}
\eea
Because of the latter identity, it is possible to extend the definition of a complex segment to matrices:
for two given angles $\alpha$ and $\beta$, with $-\pi\le\alpha<\beta\le\pi$
and $\beta-\alpha\le\pi$,
\be
\cS_{\alpha,\beta}(\M_d(\C)) := \{C\in\M_d(\C): e^{-i\alpha}C\in \cI^+ \mbox{ and } e^{-i\beta}C\in \cI^- \}.\label{eq:acutematrix}
\ee
Again, we shall drop the arguments and write $\cS_{\alpha,\beta}$ if the context is clear.
It has to be kept in mind, however, that there exist matrices that
are not in $\cS_{\alpha,\beta}$ for any value of $\alpha,\beta$.
For example, for the matrix $X = \twomat{1}{i}{i}{-1}$, it is easy to show that $\im e^{-i\alpha}X$ has eigenvalues $\pm1$, hence is indefinite, for any
value of the angle $\alpha$.

In the remainder of this section we will show that complex matrix segments satisfy most (but not all) properties that scalar segments
satisfy.
We start with some basic properties that are very easy to prove.
\begin{lemma}\label{lem:basic}
\begin{enumerate}
\item If the angle $\phi$ is such that $-\pi \le \alpha+\phi$ and $\beta+\phi\le \pi$,
then $C\in \cS_{\alpha,\beta}$ implies $e^{i\phi} C\in \cS_{\alpha+\phi,\beta+\phi}$.
\item $C\in \cS_{\alpha,\beta}$ implies $C^* \in \cS_{-\beta,-\alpha}$.
\item
For any invertible $K$, if $C\in \cS_{\alpha,\beta}$, then so is $KCK^*$.
This follows from Lemma \ref{lem:conjug}. Invertibility of $K$
is needed to ensure that conjugation of a positive definite matrix with $K$ remains positive definite (and does not become positive semidefinite).
\item If $C\in \cS_{\alpha,\beta}$ and $a>0$ then $aC\in \cS_{\alpha,\beta}$.
If $a<0$, however, then $aC\in \cS_{\alpha-\pi,\beta-\pi}$ if $0\le\alpha<\beta\le\pi$, and $aC\in \cS_{\alpha+\pi,\beta+\pi}$ if
$-\pi\le\alpha<\beta\le0$.
If $\alpha<0<\beta$ then the segment containing $aC$ necessarily has to straddle the cut, and this will not be considered.
\end{enumerate}
\end{lemma}

The next lemma shows that complex matrix segments behave in the expected way as regards interval inclusion.
\begin{lemma}\label{lem:inclusion}
Let $-\pi\le\alpha_1\le\alpha_2<\beta_2\le \beta_1\le\pi$ and $\beta_1-\alpha_1\le\pi$.
Then $\cS_{\alpha_2,\beta_2}(\M)\subset\cS_{\alpha_1,\beta_1}(\M)$.
\end{lemma}
\textit{Proof.}
We will only prove a special case; the general case follows easily using Lemma \ref{lem:basic}.
Let $-\pi/2\le\alpha<\beta\le \pi/2$. We will show that any matrix in $\cS_{\alpha,\beta}(\M)$ is also in $\cR^+(\M)$.

Let $\delta_1=\alpha-(-\pi/2)$ and $\delta_2=\pi/2-\beta$. Then $0\le \delta_1,\delta_2\le\pi$ and $0\le \delta_1+\delta_2<\pi$.
If $\delta_1=\delta_2=0$ there is nothing to prove, so let us henceforth assume that $\delta_1+\delta_2>0$.

Let $C=A+iB\in\cS_{\alpha,\beta}$. Thus both $\exp(-i\delta_1)C$ and $\exp(i\delta_2)C$ are in $\cR^+$, i.e.\ have positive real part.
Explicitly:
\[
\cos\delta_1 A > -\sin\delta_1 B \mbox{ and }
\cos\delta_2 A > \sin\delta_2 B.
\]
Noting that $\sin\delta_1$ and $\sin\delta_2$ are non-negative and not both zero, we multiply the former inequality by $\sin\delta_2$,
the latter inequality by $\sin\delta_1$ and then add both resulting inequalities.
This yields
\[
(\cos\delta_1 \sin\delta_2+\cos\delta_2\sin\delta_1)A> 0,
\]
which immediately implies $A>0$ since $\cos\delta_1 \sin\delta_2+\cos\delta_2\sin\delta_1 = \sin(\delta_1+\delta_2)>0$.
This shows that, indeed, $C$ has positive real part.
\qed

Next, we prove the very useful property that if a matrix is in a certain complex segment then its spectrum is in that same segment.
The converse of this lemma is not true: as noted before, there exist matrices that are not even contained in any complex matrix segment.
However, the converse is true for diagonal matrices (obviously), and therefore also for normal matrices, because these can be diagonalised
by (unitary) conjugation and by Lemma \ref{lem:basic} matrix segments are invariant under conjugation.
\begin{lemma}
Let two angles $\alpha$ and $\beta$ be given, with $-\pi\le\alpha<\beta\le\pi$
and $\beta-\alpha\le\pi$.
Then $C\in \cS_{\alpha,\beta}$ implies $\Sp C\subset \cS_{\alpha,\beta}$.
\end{lemma}
\textit{Proof.}
Let us first show this for $\cS_{0,\pi} = \cI^+$.
Clearly, if $\im C>0$ and $\im z\le0$, then $\im(C-z)>0$, which by Lemma \ref{lem:inv}
implies that $C-z$ is invertible. Hence, $C-z$ is invertible for all $z$ on the real axis or in the lower half-plane.
As the spectrum of $C$ consists, by definition, of those points $z$ where $C-z$ is \emph{not} invertible, this shows that $\Sp C$
lies entirely in the upper half-plane.

To show the statement in full generality, we apply this reasoning to $e^{-i\alpha}C$ and $-e^{-i\beta}C$. Noting that
$\Sp(e^{-i\alpha}C) = e^{-i\alpha}\Sp C$, we then get
$e^{-i\alpha}\Sp C \in \cI^+$ and $e^{-i\beta}\Sp C \in \cI^-$. Hence, $\Sp C\in \cS_{\alpha,\beta}$.
\qed

The main usefulness of complex matrix segments in the present context
is due to the following lemma, which generalizes Lemma 2 in \cite{epstein}.
\begin{lemma}\label{lem:specprod}
Consider two complex segments $\cS_{\alpha_i,\beta_i}$, for $i=1,2$,
(thus, the angles satisfy $-\pi\le \alpha_i<\beta_i\le \pi$ and $\beta_i-\alpha_i\le\pi$),
such that $-\pi\le\alpha_1+\alpha_2<\beta_1+\beta_2\le\pi$
and $(\beta_1+\beta_2) - (\alpha_1+\alpha_2) \le\pi$.
Let $A_i\in\cS_{\alpha_i,\beta_i}(\M_d(\C))$ ($i=1,2$).
%Then $\Sp(A_1A_2)$ lies in the complex segment $\cS_{\alpha_1+\alpha_2,\beta_1+\beta_2}(\C)$.
Then the eigenvalues of $A_1A_2$ have complex arguments in the interval $(\alpha_1+\alpha_2,\beta_1+\beta_2)$.
\end{lemma}
%[TODO: Mention that here we allow $(\beta_1+\beta_2)-(\alpha_1+\alpha_2)$ to be larger than $\pi$.]

Note that no such conclusion can be drawn from statements about the spectra of $A_1$ and $A_2$.

The proof we give below is completely different from Epstein's and much more elementary, as it does not involve deep analytical results
such as Bochner's tube theorem.

\textit{Proof.}
By Lemma \ref{lem:inclusion}, the fact that $A_i$ lies in a complex segment, $\cS_{\alpha_i,\beta_i}(\M_d(\C))$,
implies that there exist angles $\gamma_i$
such that $e^{i\gamma_i}A_i$ is in $\cR^+$. This is so if $\gamma_i\in [-\alpha_i-\pi/2,-\beta_i+\pi/2]$.
Then, by Lemma \ref{lem:spAB}, no eigenvalue of
$e^{i(\gamma_1+\gamma_2)}A_1A_2$ lies on the cut $\R^-$ if $\gamma_i$ is in the stated interval, corresponding to
$\gamma_1+\gamma_2$ lying in the interval $[-\pi-(\alpha_1+\alpha_2),\pi-(\beta_1+\beta_2)]$.
Thus, no eigenvalue of $A_1A_2$ has complex argument $-\pi-(\gamma_1+\gamma_2)$, nor $+\pi-(\gamma_1+\gamma_2)$.

By letting $\gamma_1$ and $\gamma_2$ assume all values in the stated intervals,
we find that no eigenvalue of $A_1A_2$ has complex argument less than or equal to $\alpha_1+\alpha_2$ or more than or equal to $\beta_1+\beta_2$.
In other words, they lie inside the open interval $(\alpha_1+\alpha_2,\beta_1+\beta_2)$. The conditions of
the Lemma ensure that $\alpha_1+\alpha_2$ and $\beta_1+\beta_2$ remain in the interval $(-\pi,\pi)$.
\qed
%%%%%%
\subsubsection{Fractional matrix powers}\label{sec:power}
As Epstein's method relies heavily on complex analysis,
we will need a definition of the fractional power $A^p$ (with $0<p<1$) that also applies
to general complex matrices, not just the positive definite ones.

For $0<p<1$, the complex function $z\mapsto z^p$ is defined in the cut plane
$\C\setminus \R^- = \{z: \im z\neq 0 \mbox{ or } \re z>0\}$ by
$\exp(p(\log|z| + i \arg z))$, for $\arg z \in (-\pi, \pi)$.
One can show that $z^p$
%is in the Pick class $P(0,+\infty)$  (\cite{donoghue}, Chapter II), which leads to
has the following integral representation:
\bea
z^p
&=& \frac{\sin p\pi}{\pi} \int_0^\infty dt\; t^{p-1} \frac{z}{t+z} \nonumber \\
&=& \frac{\sin p\pi}{\pi} \int_0^\infty dt\; t^p \left(\frac{1}{t}-\frac{1}{t+z}\right).\label{eq:fracscalar}
\eea
This representation can be used to extend the definition of fractional powers to matrices in the obvious way.
For any matrix $C$ with spectrum $\Sp C \subset \C\setminus\R^-$,
\be
C^p = \frac{\sin p\pi}{\pi} \int_0^\infty dt\; t^p \left(\frac{1}{t}-(t+C)^{-1}\right).\label{eq:fracmatrix}
\ee
This definition ensures that for any holomorphic operator-valued function $C(z)$
the function $C(z)^p$ is again holomorphic whenever the spectrum of $C(z)$ remains in $\C\setminus\R^-$.

Using this integral, we can express $\im C^p$ as an integral as well:
\bea
\im C^p &=& \frac{\sin p\pi}{\pi} \int_0^\infty dt\; t^p \im\left(\frac{1}{t}-(t+C)^{-1}\right)\nonumber \\
&=& \frac{\sin p\pi}{\pi} \int_0^\infty dt\; t^p \im(-(t+C)^{-1}).\label{eq:fracmatrixim}
\eea

The effect of a fractional power on the complex matrix segments is as follows:
\begin{lemma}\label{lem:powersegment}
Let $C\in \cS_{\alpha,\beta}(\M_d(\C))$. If $0<p<1$ then $C^p\in \cS_{p\alpha,p\beta}(\M_d(\C))$.
If $-1<p<0$, then $C^p\in \cS_{p\beta,p\alpha}(\M_d(\C))$.
\end{lemma}
\textit{Proof.}
The condition $|p|<1$ ensures that the angles of the complex segment $\cS_{p\alpha,p\beta}(\C)$ satisfy
the condition $p\beta-p\alpha\le\pi$. Consider first the case $0<p<1$.
Again we first consider the case $C\in \cI^+$.
Then $\im C>0$ and
by Lemma \ref{lem:invert},
\be
\im(-C^{-1}) = C^{-1} \im(C) C^{-*} >0.\label{eq:lempowersegment}
\ee
Combining this with the integral expression (\ref{eq:fracmatrixim}) we find that $\im C^p>0$.
To prove the general statement,
we apply this to the matrices $e^{-i\alpha}C$ and $-e^{-i\beta}C$ and get
$\im (e^{-i\alpha}C)^p = \im e^{-ip\alpha}C^p>0$ and $\im e^{-ip\beta}C^p<0$, which shows that
$C^p\in \cS_{p\alpha,p\beta}$.

For the case $-1<p<0$, note that $C^p$ is just $(C^{-1})^{|p|}$ and that if $C\in\cI^+$, then by (\ref{eq:lempowersegment})
$-C^{-1}\in\cI^+$ as well.
Applying the preceding argument to $-C^{-1}=e^{-i\pi}C^{-1}$ yields the statement for $-1<p<0$.
\qed

The restriction $|p|\le1$ is essential. For $p>1$ it might seem plausible that if $C\in \cS_{\alpha/p,\beta/p}$,
with $\beta-\alpha\le\pi$, then $C^p\in \cS_{\alpha,\beta}$;
however, this is not true. Taking $p=2$, for example, then $C^2\in \cS_{-\pi/2,\pi/2}=\cR^+$ iff $\re C^2>0$,
which amounts to $(\re C)^2\ge(\im C)^2$.
In contrast, $C\in\cS_{-\pi/4,\pi/4}$ iff $\re C\ge \im C$ and $\re C\ge -\im C$, which is not sufficient to ensure
$(\re C)^2\ge(\im C)^2$ (as the function $x^2$ is not operator monotone).

%%%%%%%%%%
\subsection{Proof that $F$ is a Pick function}
After these preliminaries, we can now turn to the actual proof that $F(t)$ is a Pick function.
Still keeping $A_1,A_2>0$ fixed, we extend the definition of $F$ to the complex plane.
For $\zeta:=x+iy$ in appropriate subsets of $\C$ (to be determined below) we define
\be
F(\zeta) := f_{p,q}(A_1+\zeta A_2) = f_{p,q}(A_1+x A_2 + iy A_2),
\ee
where the domain of $f_{p,q}$ has been extended to complex matrices.
We will show that
$F$ is in the class $P(\tau,+\infty)$, where $\tau$ is
a negative number depending on $A_1$ and $A_2$.

%%%%%%%%%%%%%
\subsubsection{Domain of holomorphy of $F$}
%[TODO: we only need $\cR^+$ and $\cI^+$, so this could be shortened ???]

First we need to establish the domain of holomorphy of $F$.
For the purposes of the proof, it is sufficient to extend the domain of $f_{p,q}$ to the set of matrices in $\M_d(\C)$
that are contained in one (or more) of the half-planes $\cS_{\theta-\pi/2,\theta+\pi/2}$ for some $\theta\in[-\pi/2,\pi/2]$.
%Thus,
%\beas
%D &:=& \bigcup_{\theta\in[-\pi/2,\pi/2]} \cS_{\theta-\pi/2,\theta+\pi/2}(\M_d(\C)) \\
%&=& \{A\in \M_d(\C): \re e^{-i\theta} A>0, \mbox{ for some }\theta\in[-\pi/2,\pi/2]\}.
%\eeas
%Note that in the scalar case
%$$
%\bigcup_{\theta\in[-\pi/2,\pi/2]} \cS_{\theta-\pi/2,\theta+\pi/2}(\C)
%$$
%is just the cut complex plane $\C\setminus\R^-$.
%
%\bigskip

The function $F(\zeta)$ is therefore well-defined when $A_1+\zeta A_2 = (A_1+xA_2)+iyA_2$ is in one of these half-planes.
One sufficient condition for this is that $y\neq 0$. Indeed, if $y>0$ then $\im(A_1+\zeta A_2) = \re(-i(A_1+\zeta A_2))>0$,
so that $A_1+\zeta A_2\in \cI^+$; if $y<0$ then $\im(A_1+\zeta A_2) = -\re(i(A_1+\zeta A_2))<0$,
and $A_1+\zeta A_2\in \cI^-$.
Another sufficient condition is $A_1+xA_2>0$, and this is so if $x>\tau$, where $\tau$ is the largest value of $x$ for which $A_1+xA_2\ge0$.
Clearly, $\tau\le0$.
Thus, it suffices to let the domain of $F(\zeta)$ be the set $\C\setminus (-\infty,\tau]$.

Since $F(\zeta)$ is a composition of holomorphic functions, it is holomorphic itself in this domain.
Furthermore, note that $F$ satisfies the reflection identity: $F(\overline{z}) = \overline{F(z)}$.
This is easy to verify from the definition of $F$, but it also
follows from the Schwarz reflection principle: since $F$ is real and continuous on the real line (excluding the cut $[-\infty,\tau]$),
so that $F$ can be continued analytically from the upper half-plane through the real interval $(\tau,+\infty)$ to the lower
half-plane by reflection.
\subsubsection{$F$ is a Pick function}
To show that $F$ is a Pick function, we need to show that $\im f_{p,q}(B_1+iB_2)>0$ whenever $B_1=B_1^*$ and $B_2>0$
(here $B_1 = A_1+xA_2$ and $B_2=yA_2$).
From the reflection identity we then also have that $\im f_{p,q}(B_1+iB_2)<0$ whenever $B_2<0$;
however, we do not actually need to invoke this identity.
By homogeneity of $f_{p,q}$, we have $f_{p,q}(A) = f_{p,q}(-iA)/(-i) = f_{p,q}(iA)/i$, so that both conditions
are equivalent to the one condition $\re f_{p,q}(A)>0$ whenever $A\in\cR^+$.

The conditions on $p$ and $q$ are that $p,q\in(0,1]$. Thus, if $A\in\cR^+ = \cS_{-\pi/2,\pi/2}$, then
by Lemma \ref{lem:basic} and Lemma \ref{lem:powersegment}, $A^p$ and $K A^q K^*$ are in the complex segments
$\cS_{-p\pi/2,p\pi/2}$ and $\cS_{-q\pi/2,q\pi/2}$, respectively.
Lemma \ref{lem:specprod} then shows that the eigenvalues of $A^p K A^q K^*$ have complex argument in the open interval
\[
(-(p+q)\pi/2,(p+q)\pi/2).
\]
On taking the $1/(p+q)$th power, this yields
$$
\Sp((A^p K A^q K^*)^{1/(p+q)}) = \Sp(A^p K A^q K^*)^{1/(p+q)}\subset \cS_{-\pi/2,\pi/2}(\C) = \cR^+,
$$
and, along with the analyticity properties previously discussed, this proves that $F$ is in the Pick class $P(\tau,+\infty)$.

%%%%%%%%%%%%%%%%%%%%%%%%%%%%%%%%%%%%%%%%%%%%%%%%%%%%%%%%%%%%%%%%%%%%%%%%%%%%%%%%%%%%%%%%%%%%%%%%%%%%%%%%%%%%%%%%%%%%%%%%%%%%%%%%%%%%%%%%%%%%%%
\section{Discussion}\label{discuss}
In this paper we studied a two-parameter family of relative entropies, which we call the $\alpha$-$z$-R\'enyi relative entropies ($\alpha$-$z$-RRE),
from which all other known relative entropies (or divergences) can be derived.
This family provides a unifying framework for the analysis of properties of the different
relative entropies arising in quantum information theory,
such as the quantum relative entropy, the $\alpha$-quantum R\'enyi divergences ($\alpha$-QRD),
and the $\alpha$-quantum R\'enyi relative entropies. We have shown that the $\alpha$-$z$-RRE satisfies the data-processing inequality (DPI)
for suitable values of the parameters $\alpha$ and $z$.

The $\alpha$-QRD (or sandwiched  R\'enyi relative entropy), which is a special case of the $\alpha$-$z$-RRE,
has been the focus of much research of late.
We have studied another special case of the $\alpha$-$z$-RRE, which we denote as  $\widehat D_\alpha$ (and informally call the
\emph{reverse sandwiched  R\'enyi relative entropy}). It satisfies DPI for $\alpha \le 1/2$, and we
obtain an interesting closed expression for it in the limit $\alpha \to 1$.

Our analysis leads to some interesting open questions:
(i) Does the $\alpha$-$z$-RRE satisfy DPI in the orange regions of the $\alpha$-$z$-plane of Figure~1?
In other words, is the trace functional of the $\alpha$-$z$-RRE convex in the orange regions of Figure~2?
(ii) Operational relevance of the $\alpha$-QRD  for $\alpha\ge1$ has been established in
quantum hypothesis testing \cite{milan2}, and in the context of the second laws of quantum thermodynamics \cite{thermo}.
Does $\widehat{D}_\alpha$ also have operational interpretations in quantum information theory (for
$0 \le \alpha \le 1/2$) (other than those arising through the symmetry relation (\ref{eq:hatsymm}))?

%%%%%%%%%%%%%%%%%%%%%%%%%%%%%%%%%%%%%%%%%%%%%%%%%%%%%%%%%%%%%%%%%%%%%%%%%%%%%%%%%%%%%%%%%%%%%%%%%%%%%%%%%%%%%%%%%%%%%%%%%%%%%%%%%%%%%%%%%%%%%%
\section*{Acknowledgments}
We are grateful to William Wootters for
suggesting the idea of looking at ``reverse sandwiching''.
This project was initiated during the program ``Mathematical Challenges in Quantum Information''
at the Isaac Newton Institute, Cambridge, UK, and was supported
in part by an Odysseus grant from the Flemish Fund for Scientific Research (FWO).
We thank the anonymous referee for the many comments that have greatly improved the clarity of the paper.
%%%%%%%%%%%%%%%%%%%%%%%%%%%%%%%%%%%%%%%%%%%%%%%%%%%%%%%%%%%%%%%%%%%%%%%%%%%%%%%%%%%%%%%%%%%%%%%%%%%%%%%%%%%%%%%%%%%%%%%%%%%%%%%%%%%%%%%%%%%%%%

\end{document}